\documentclass[useAMS,usenatbib,psfig]{mn2e}

\usepackage{xspace}
\usepackage{graphicx}

\newcommand {\apj} {ApJ}
\newcommand {\apjl} {ApJL}
\newcommand {\apjs} {ApJS}
\newcommand {\mnras} {MNRAS}

\newcommand {\aap} {A\&A}
\newcommand {\aj} {AJ}

\newcommand {\araa} {ARA\&A}

\newcommand {\pasp} {PASP}

\newcommand {\etal} {et~al.~}
\def \spose#1{\hbox  to 0pt{#1\hss}}  
\newcommand {\lta} {\mathrel{\spose{\lower 3pt\hbox{$\sim$}}\raise  2.0pt\hbox{$<$}}}
\newcommand {\gta} {\mathrel{\spose{\lower  3pt\hbox{$\sim$}}\raise 2.0pt\hbox{$>$}}}

\newcommand {\ha}  {\ifmmode H\alpha \else H$\alpha $ \fi} 

\def\Sref#1{Section~\ref{#1}\xspace}

\newcommand {\kms} {\ifmmode  \,\rm km\,s^{-1} \else $\,\rm km\,s^{-1}  $ \fi }
\newcommand {\kpc} {\ifmmode  {\rm kpc}  \else ${\rm  kpc}$ \fi  }  
\newcommand {\pc} {\ifmmode  {\rm pc}  \else ${\rm pc}$ \fi  }  
\newcommand {\Msun} {\ifmmode {\rm M_{\odot}} \else ${\rm M_{\odot}}$ \fi} 
\newcommand {\Zsun} {\ifmmode {\rm Z_{\odot}} \else ${\rm Z_{\odot}}$ \fi} 
\newcommand {\yr} {\ifmmode yr^{-1} \else $yr^{-1}$ \fi} 
\newcommand {\hMsun} {\ifmmode h^{-1}\,\rm M_{\odot} \else $h^{-1}\,\rm M_{\odot}$ \fi}

\def\zd{z_{\rm d}}
\def\zs{z_{\rm s}}

\def\q3{q_{3}}

\def\hst{{\it HST}\xspace}

\def\galfit{{\sc galfit}\xspace}

\def\SPASMOID{{\sc SPASMOID}\xspace}

\def\sigmasdss{$\sigma_{\rm SDSS}$}

\def\NSWELLS{27}

\def\NSWELLSA{8}
\def\NSWELLSB{1}
\def\NSWELLSC{6}
\def\NSWELLSX{4}
\def\NSSLACS{10}
\def\NSESLACS{8}

\def\NTOTALA{16}
\usepackage[usenames]{color}

\def\uvic{Dept. of Physics and Astronomy, 
  University of Victoria, Victoria, BC, V8P 5C2, Canada}
\def\lick{UCO/Lick Observatory, Department of Astronomy and Astrophysics, 
  University of California, Santa Cruz, CA 95064, USA}
\def\ucsb{Dept. of Physics, University of California, 
  Santa Barbara, CA 93106, USA}
\def\kipac{Kavli Institue for Particle Astrophysics and Cosmology, 
  P.O. Box 20450, MS29, Stanford, CA 94309, USA}
\def\utah{Department of Physics and Astronomy, University of Utah, 
  Salt Lake City, UT 84112, USA}
\def\kapteyn{Kapteyn Astronomical Institute, University of Groningen, 
  P.O.Box 800, 9700 AV Groningen, The Netherlands}
\def\oxford{Department of Physics, University of Oxford, 
  Keble Road, Oxford, OX1 3RH, UK}

\def\treuemail{\tt tt@physics.ucsb.edu}

\def\packard{Packard Research Fellow}
\def\cita{CITA National Fellow}

\newcommand\nodata{ ~$\cdots$~ }%

\title[The SWELLS Survey. I.]
{The SWELLS Survey. I. A large spectroscopically selected sample of edge-on late-type lens galaxies.}
    
\author[Treu \etal]{%
  Tommaso~Treu$^{1}$\thanks{\treuemail}\thanks{\packard},
  Aaron~A.~Dutton$^{2,3}$\thanks{\cita},
  Matthew~W.~Auger$^{1}$,
  Philip~J.~Marshall$^{1,4,5}$,
\newauthor{%
  Adam~S.~Bolton$^{6}$,
  Brendon~J.~Brewer$^{1}$,
  David~C.~Koo$^{3}$,
  L\`eon~V.~E.~Koopmans$^{7}$}
  \medskip\\
  $^1$\ucsb\\
  $^2$\uvic\\
  $^3$\lick\\
  $^4$\kipac\\
  $^5$\oxford\\
  $^6$\utah\\
  $^7$\kapteyn
}

\begin{document}
             
\date{submitted to MNRAS}
             
\pagerange{\pageref{firstpage}--\pageref{lastpage}}\pubyear{2010}

\maketitle           

\label{firstpage}


\begin{abstract}
The relative contribution of baryons and dark matter to the inner
regions of spiral galaxies provides critical clues to their formation
and evolution, but it is generally difficult to determine. For spiral
galaxies that are strong gravitational lenses, however, the
combination of lensing and kinematic observations can be used to break
the disk-halo degeneracy. In turn, such data constrain fundamental
parameters such as i) the mass density profile slope and axis ratio of
the dark matter halo, and by comparison with dark matter-only
numerical simulations the modifications imposed by baryons; ii) the
mass in stars and therefore the overall star formation efficiency, and
the amount of feedback; iii) by comparison with stellar population
synthesis models, the normalization of the stellar initial mass
function. In this first paper of a series, we present a sample of 16
secure, 1 probable, and 6 possible strong lensing spiral galaxies, for
which multi-band high-resolution images and rotation curves were
obtained using the Hubble Space Telescope and Keck-II Telescope as
part of the Sloan WFC Edge-on Late-type Lens Survey (SWELLS). The
sample includes 8 newly discovered secure systems. We characterize the
sample of deflector galaxies in terms of their morphologies,
structural parameters, and stellar masses.  We find that the SWELLS
sample of secure lenses spans a broad range of morphologies (from
lenticular to late-type spiral), spectral types (quantified by
H$\alpha$ emission), and bulge to total stellar mass ratio
(0.22-0.85), while being limited to M$_*>10^{10.5}$ M$_\odot$. The
SWELLS sample is thus well-suited for exploring the relationship
between dark and luminous matter in a broad range of galaxies. We find
that the deflector galaxies obey the same size-mass relation as that
of a comparison sample of elongated non-lens galaxies selected from
the SDSS survey. We conclude that the SWELLS sample is consistent with
being representative of the overall population of high-mass
high-inclination disky galaxies.
\end{abstract}



\setcounter{footnote}{1}


\section{Introduction}

The discovery of extended flat rotation curves in the outer parts of
disk galaxies three decades ago \citep{Bos78,RTF78} was decisive in
ushering the paradigm shift that led to the now standard cosmological
model dominated by cold dark matter (CDM) and dark energy.  On
cosmological (i.e.\ linear) scales there is excellent agreement
between the predictions of the standard $\Lambda$CDM model and
observations of the CMB, type Ia Supernovae, weak-lensing, and galaxy
clustering \citep[e.g.\ ][]{Spe++07}.  On galactic and sub-galactic
(i.e. non-linear) scales, however, there are a number of apparent
discrepancies between the predictions \citep[e.g.\ ][] {NFW97,Bul++01}
and observations \citep[e.g.\ ][]{Moo++99,Kly++99,deBlo++01,
Swa++03,Dut++07,San++08,New++11} of the structure and mass function of
dark matter haloes. Whether these problems reflect an incomplete
understanding of galaxy formation, or signal an unresolvable problem
for the standard paradigm, remains to be determined. Measuring the
density profiles of dark matter halos on galaxy scales is thus a
crucial test for the standard paradigm of galaxy formation, offering
great potential for discovery.

In most luminous galaxies the baryons (i.e.\ stars and cold gas)
always make a non-negligible contribution to the gravitational
potential, and due to the so-called {\it disk-halo degeneracy}
\citep{A+S86} the distribution of baryons and dark matter in galaxies
is poorly constrained, even when high spatial resolution rotation
curves are available \citep[e.g.\ ][]{vdB+S01,Dut++05}.

\begin{table*}
\begin{center}
\scriptsize\begin{tabular}{cccccccccc}
\hline
\hline 
ID    & RA   & DEC  & $\zd$ & $\zs$& Grade & \sigmasdss & H$\alpha$     	          & $\left(b/a\right)_{\rm SDSS}$ & Ref \\
      &      &      &       &      &       & (\kms)     & (10$^{-17}$ergs$^{-1}$cm$^{-2}$) &                  &     \\
(1)   & (2)  & (3)  & (4)   & (5)  & (6)   & (7)        & (8)                            & (9)             & (10)\\
\hline
J0820+4847 & 125.05363 & 48.79364 & 0.131 & 0.634 & A & 168$\pm$9  & \nodata  & 0.32 & 3 \\ 
J0821+1025 & 125.48858 & 10.43226 & 0.094 & 0.657 & C & 144$\pm$12 & 58.4$\pm$3.2 & 0.22 & 3 \\ 
J0841+3824 & 130.37004 & 38.40381 & 0.116 & 0.657 & A & 216$\pm$8  & 55.5$\pm$7.2 & 0.55 & 2 \\ 
J0915+4211 & 138.81787 & 42.19800 & 0.078 & 0.790 & A & 176$\pm$7  & 33.6$\pm$6.4 & 0.46 & 3 \\ 
J0955+0101 & 148.83217 & 1.02901  & 0.111 & 0.316 & A & 196$\pm$13 & 27.5$\pm$5.8 & 0.51 & 2 \\ 
J1015+1750 & 153.76331 & 17.84731 & 0.129 & 0.376 & X & 169$\pm$17 & 75.6$\pm$3.2 & 0.29 & 3 \\ 
J1029+0420 & 157.34560 & 4.33384  & 0.104 & 0.615 & A & 208$\pm$9  & \nodata  & 0.54 & 1 \\ 
J1032+5322 & 158.14932 & 53.37636 & 0.133 & 0.329 & A & 293$\pm$15 & \nodata  & 0.57 & 2 \\ 
J1037+3517 & 159.43764 & 35.29194 & 0.122 & 0.448 & A & 223$\pm$12 & 23.4$\pm$5.2 & 0.49 & 3 \\ 
J1056+0005 & 164.13266 & 0.08470  & 0.039 & 0.315 & X & \nodata    & 34.3$\pm$1.4 & 0.40 & 3 \\ 
J1103+5322 & 165.78421 & 53.37450 & 0.158 & 0.735 & A & 201$\pm$12 & \nodata  & 0.46 & 2 \\ 
J1117+4704 & 169.39742 & 47.06873 & 0.169 & 0.405 & A & 187$\pm$10 & \nodata  & 0.43 & 3 \\ 
J1135+3720 & 173.77867 & 37.33997 & 0.162 & 0.402 & A & 213$\pm$15 & 20.0$\pm$6.1 & 0.27 & 3 \\ 
J1203+2535 & 180.98470 & 25.59697 & 0.101 & 0.856 & A & 156$\pm$7  & 13.2$\pm$3.9 & 0.32 & 3 \\ 
J1228+3743 & 187.24837 & 37.73011 & 0.040 & 0.102 & B & 180$\pm$5  &   85$\pm$15  & 0.46 & 3 \\ 
J1251$-$0208 & 192.89877 & $-$2.13477 & 0.224 & 0.784 & A & 203$\pm$22 & 41.2$\pm$5.3 & 0.51 & 1 \\ 
J1258$-$0259 & 194.62041 & $-$2.99444 & 0.111 & 0.507 & X & 152$\pm$9  & \nodata  & 0.36 & 3 \\ 
J1300+3704 & 195.14450 & 37.07192 & 0.016 & 0.160 & X & \nodata    & 111$\pm$2.0  & 0.28 & 3 \\ 
J1313+0506 & 198.36127 & 5.11589  & 0.144 & 0.338 & A & 249$\pm$16 & \nodata  & 0.51 & 2 \\ 
J1321$-$0115 & 200.31609 & $-$1.25500 & 0.108 & 0.211 & C & 189$\pm$12 & 20.4$\pm$6.2 & 0.48 & 3 \\ 
J1331+3638 & 202.91800 & 36.46999 & 0.113 & 0.254 & A & 175$\pm$8  & 24.3$\pm$5.0 & 0.56 & 3 \\ 
J1410+0205 & 212.57333 & 2.09115  & 0.127 & 0.734 & C & 171$\pm$13 & \nodata  & 0.43 & 3 \\ 
J1521+5805 & 230.34937 & 58.09747 & 0.204 & 0.486 & C & 182$\pm$15 & 29.6$\pm$4.7 & 0.49 & 3 \\
J1556+3446 & 239.05722 & 34.77497 & 0.073 & 0.598 & C & 157$\pm$5  & 30.4$\pm$4.8 & 0.29 & 3 \\ 
J1703+2451 & 255.92278 & 24.86111 & 0.063 & 0.637 & A & 165$\pm$8  &476.4$\pm$6.0 & 0.28 & 3 \\ 
J2141$-$0001 & 325.47781 & $-$0.02008 & 0.138 & 0.713 & A & 183$\pm$13 & 21.7$\pm$3.9 & 0.30 & 2 \\ 
J2210$-$0934 & 332.52393 & $-$9.57112 & 0.083 & 1.158 & C & 133$\pm$10 & \nodata  & 0.25 & 3 \\ 
\hline \hline
\end{tabular}
\caption{Summary of basic properties of the SWELLS sample of spiral
lens galaxies.  Col.~1 lists the lens ID; Cols.~2 and~3 the
coordinates (J2000); Cols~4 and~5 give deflector and source redshifts;
Col.~6 the lens grade (A=secure, B=probable, C=possible, X=not a lens)
according to the definition of \citep{Bol++06}; Col.~7 the velocity
dispersion within the spectroscopic fiber as measured by SDSS; Col.~8
the H$\alpha$ flux as measured by SDSS spectroscopy based on the subtracted model spectrum; Col~9. the axis
ratio as measured by SDSS; Col.~10 the reference for the discovery of
the lens (1=Bolton et al.\ 2006, 2=Bolton et al.\ 2008, 3=this
paper). \label{tab:sample}}
\end{center}
\end{table*}

A potentially powerful method for breaking the disk-halo degeneracy is
to combine galaxy kinematics with strong gravitational lensing
\citep{Mal++00}. This method makes use of the fact that kinematics
measures mass enclosed within a sphere, while strong lensing measures
projected mass
\citep[e.g.\ ][and references therein]{Tre10}.  The power of strong
gravitational lensing and dynamics to break the disk-halo degeneracy
has, to date, not been fully realized, due to the scarcity of known
spiral galaxy gravitational lenses.  Up until a few years ago, only a
handful of spiral galaxy lenses with suitable inclinations to enable
rotation curve measurements were known: Q2237$+$0305
\citep{Huc++85}; B1600$+$434 \citep{J+H97,Koo++98}; PMN~J2004$-$1349
\citep{WHS03}; CXOCY~J220132.8$-$320144 \citep{Cas++06}. However most
of these systems are doubly imaged QSOs, which provide minimal
constraints on the projected mass distribution. Q2237$+$0305 is a
quadruply imaged QSO, which gives more robust constraints, but since
the Einstein radius is five times smaller than the half light size of
the galaxy, the lensing is mostly sensitive to the bulge mass
\citep{Tro++10}. Recently, extensive searches based on large parent
surveys have started to find gravitational lens candidates where the
deflector is a spiral galaxy. Examples of imaging-based searches for
spiral lens galaxies include those by \citet{Syg++10},
\citet{Mar++09}, and \citet{Mor++11}, while spectroscopy-based
searches include those by \citet{Fer++09} and the SLACS Survey
(\citeauthor{Bol++06}~\citeyear{Bol++06}, \citeyear{Bol++08},
\citeauthor{Aug++09}~\citeyear{Aug++09}).  One of the main challenges
of identifying and studying strong lens systems where the deflector is
a spiral galaxy is the presence of dust. Modeling and correcting for
dust obscuration is essential to model the gravitational potential of
the deflector; this can be done by exploiting the conservation of
surface brightness of the background source in gravitational lensing.

Since its beginning in Hubble Space Telescope (HST) Cycle~13, SLACS
has discovered 85 strong galaxy-galaxy lenses (and an additional 13
high probability candidates).  Because of the strong dependency of the
lensing cross-section on central surface mass density, $\sim90$\% of
the lenses discovered by SLACS have been massive early-type galaxies
\citep{Aug++09}. However, thanks to its large overall size, SLACS was
able to discover a considerable number of previously unknown
gravitational lens systems with a deflector of spiral
morphology. \NSSLACS\ SLACS lenses have visible spiral morphology
(classification S or S0/Sa). Of these
\NSSLACS, \NSESLACS\ have disk inclinations high enough ($b/a<0.6$)
for measuring reliable rotation curves. This subset alone already more
than doubles the sample of previously known systems. In addition,
source and deflector redshifts are known by construction for the SLACS
systems, and the deflector galaxy is generally bright and at a
relatively low redshift, making it a very practical sample for
detailed follow-up. The main limitations of the SLACS-spiral sample
are that it is limited to bulge-dominated spirals, since they are the
most common in strong lensing selected samples, and that spatially
resolved kinematics is not available from the SDSS fiber-based
spectroscopy, but requires follow-up \citep{Bar++11}.

The goal of the SWELLS (Sloan WFC Edge-on Late-type Lens Survey;
WFC~=~Wide Field Camera) project is to identify, follow-up, and
analyze in a systematic fashion a large and homogeneous sample of
spiral lens galaxies. We aim to combine strong gravitational lensing
with stellar kinematics to break the ``disk-halo'' degeneracy and
address two fundamental issues.  Firstly, measurements of dark matter
halo density profiles and halo shapes will be compared to predictions
of galaxy formation models in the concordance $\Lambda$CDM cosmology.
Secondly, measurements of stellar mass-to-light ratios will be
compared to those predicted from stellar population models, thus
placing constraints on the stellar initial mass function
\citep{Tre++10,Aug++10,Spi++11}.

The sample consists of high-inclination spiral lens galaxies, drawn
from both the SLACS Survey and a new dedicated SDSS/\hst search. For
each object in its target list, the SWELLS survey is measuring high
quality rotation and velocity dispersion profiles using long-slit
spectroscopy at the Keck Telescopes. In addition, high resolution
images in multiple bands, including the infrared with Keck, are being
obtained to model and correct dust obscuration, identify and model
multiple images of the background source, and infer stellar masses of
bulge and disk from stellar population synthesis models.

In this first paper of the series we present an overview of the survey
and the lens sample, and discuss the lensing galaxies as spiral
galaxies to identify potential selection effects that must be taken
into account when interpreting our results. Papers of this series will
present detailed analyses of the individual systems illustrating the
methodology and challenges of a joint lensing and dynamical study of
spiral galaxies \citep[][hereafter paper II]{Dut++11b} and a library
of gravitational lens models (Brewer et al. 2011, in preparation).

This paper is organized as follows. Our strategy and sample selection
algorithm is described in~\Sref{sec:sample}.  The data are described in
\Sref{sec:data}. In~\Sref{sec:sampprop} we present lensing
classification, morphological and structural properties and stellar masses for
each system. In~\Sref{sec:swellsspirals} we discuss the lensing galaxies as
spiral galaxies, by comparing them to a control sample drawn from the SDSS
survey. We discuss our findings and the lessons learned with an eye to future
searches for strong lensing spirals in \Sref{sec:disc}. A brief summary is
given in~\Sref{sec:summary}. Throughout this paper, and the rest of the
SWELLS series, magnitudes are
given in the AB system \citep{Oke74} and we adopt standard ``concordance''
cosmological parameters, i.e. $h=0.7$, $\Omega_m=0.3$ and
$\Omega_\Lambda=0.7$, where the symbols indicate the Hubble Constant in units
of 100 km s$^{-1}$ Mpc$^{-1}$ and the matter and dark energy density of the
Universe in units of the critical density \citep[e.g.\ ][]{Kom++09}.


\section{Sample Selection}
\label{sec:sample}

The SWELLS sample is composed of two subsamples. The first is a
subset of the spiral lens galaxies identified by the SLACS Survey
\citep{Bol++06,Bol++08,Aug++09}, suitable for kinematic follow-up due
to their high inclination, as determined by the observed axis ratio. The
second subsample consists of new spiral lens galaxies, identified from
the SDSS-database and confirmed with \hst imaging in the following
manner. The strategy is similar to that adopted for the original SLACS
sample, optimised to identify spiral lens galaxies.

First we looked for SDSS spectra that contain two sets of lines at two
different redshifts, indicating a foreground and background object in
the same 3 arcsecond-diameter solid angle covered by the fiber as
detailed in the Appendix.  Out of almost a million spectra, our
spectroscopic search algorithm found more than 200 new
high-probability lens candidates.  Second, SDSS images of the lens
candidates were visually inspected by two of us (TT and AAD) to
identify those spiral galaxies sufficiently inclined for rotation
curve measurements, and to reject obvious failures or
contaminants. Third, we estimated the strong lensing probability
according to the following procedure. Based on our previous experience
\citep{Bol++08}, the lens confirmation rate is a well known and
monotonically increasing function of Einstein radius (effectively,
proportional to the fraction of the solid angle of the fiber that is
contained within the critical lines, plus seeing and other
observational effects). The Einstein radius was estimated from the
velocity dispersion of the lens and the usual ratio of angular
diameter distances, assuming that the stellar velocity dispersion
equals the normalisation of the best fitting singular isothermal
ellipsoid \citep{Tre++06,Bol++08}. Since velocity dispersions are not
always available or reliable, especially for the more disk-dominated
systems with emission lines, we used total luminosity as a proxy for
stellar velocity dispersion, after calibrating the correlation on the
rest of the sample. We expect this to be a conservative estimate,
since edge-on systems are more effective lenses due to their higher
projected mass density \citep{MFP97,K+K98,BMM99}.  Fourth, to optimise
use of telescope time, we excluded all targets with estimated lensing
probability below 30\%. This left a total sample of 43 candidates,
which we used to select targets for high-resolution imaging follow-up
based on observability. If more than one candidate was observable at a
given time, we gave priority to those with the highest estimated
lensing probability.  Follow-up observations and lens confirmation
rates are discussed in the next sections.

A total of 27 candidates were observed with high resolution imaging. A
summary of the properties of these SWELLS lenses and candidates is
given in Table~\ref{tab:sample}.


\section{Observations}
\label{sec:data}

\subsection{Hubble Space Telescope imaging}
\label{ssec:hst}

In the supplementary Cycle~16s we were awarded 91 orbits (of which 74
were executed) of {\it \hst} time to discover new spiral galaxy lenses
and to complete three-band optical imaging of the known SLACS spiral
lenses (GO 11978; PI~Treu). As a compromise between maximum wavelength
coverage -- needed to model and correct dust obscuration, identify
strongly lensed features, and model stellar populations -- and
efficiency, we chose to image each lens through the F450W, F606W, and
F814W filters.

As part of GO-11978, twenty new candidates (selected as described in
the previous section) were scheduled to be observed. Of those, fifteen
were observed with WFPC2 in 3 bands: F450W (two orbits), F606W (one
orbit), and F814W (one orbit), and one was observed in F450W and F606W
(two and one orbit, respectively). Due to scheduling constraints the
F814W exposure and remaining 4 targets were not observed\footnote{Three
of these remaining targets were observed with K-band laser guide star
adaptive optics (LGS-AO) imaging on the Keck II telescope as described
in~\S~\ref{ssec:keck-ao}.}. In addition, five SLACS-selected spiral
lenses (J0841+3824, J1029+0420, J1032+5322, J1103+5322, J2141-0001)
were imaged in F450W (two orbits each) and/or F606W (one orbit) to
complete three-band imaging for the sample.

All targets were placed in the centre of the WFPC2-WF3 chip, which has
a pixel scale of $0\farcs1$. The observations in each filter consisted
of four exposures dithered by fractional pixel amounts to remove
defects, identify cosmic ray hits, and to recover resolution lost to
undersampling. The raw images were reduced and combined using a custom
built pipeline based on the drizzle package, resulting in final images
with a pixel scale of 0$\farcs$05.

Additional \hst imaging is available for the SLACS subsample as listed
in Table~\ref{tab:structure}. ACS/WFCP2 and NICMOS observations are
presented in the papers by \citet{Bol++06,Bol++08} and
\citet{Aug++09}. Infrared images obtained with WFC3 through filter
F160W as part of GO-11202 (PI~Koopmans) are presented here for the
first time. As with the other instruments, a standard four-point
dither strategy was adopted, in this case to remove defects and
recover some of the information lost to the undersampling of the point
spread function (PSF) by the $0\farcs13$ pixels. The data were reduced
via a custom pipeline based on the drizzle package, using the most up
to date calibration files.

\subsection{Keck Adaptive Optics Imaging}
\label{ssec:keck-ao}

High-resolution infrared imaging is extremely valuable for dust
correction, multiple image identification and stellar mass estimation.
For these purposes, $K'$-band (2.15$\mu$m) imaging was obtained for a
subset of candidates/lenses for which a suitable bright nearby star is
available for tip-tilt correction using the Laser Guide Star Adaptive
Optics (LGSAO) system on the Keck-II 10m Telescope. Higher priority
was given to those systems for which \hst infrared imaging is not
available, although the combination of \hst-F160W and ground based
$K'$-band imaging is especially powerful.  The sensitivity of Keck AO
at longer wavelengths complements very nicely the \hst PSF stability
and sensitivity to low surface brightness features, helping in
particular with dust correction and stellar mass estimates.

All observations were performed with the NIRC2 camera in wide field
mode, yielding a pixel scale of $0\farcs04$, typical resolution of
$0\farcs10$ FWHM and strehl ratio of 0.15-0.20. The observations took
place on the following dates, during times of clear sky and good
seeing conditions suitable for AO correction: Nov 11 2009 (PI~Koo);
May 7/8 2009, August 12 2009, January 04 2010 (PI~Treu). Typical
exposure times ranged between 30 and 60 minutes, depending on
conditions and faintness of the target. Table~\ref{tab:structure}
lists the targets that were observed with AO. More details on the
observations of individual systems will be given in future papers
along with the analysis of the systems themselves. Although in general
the lensed features are rather blue and more easily identified in \hst
images, we note that two of the grade-A SWELLS lenses were observed
only with Keck-AO, and confirmed from the ground.

All NIRC2 data were reduced with a custom Python-based reduction
package described by \citet{Aug++11a}.  A sky frame and a sky flat
were created from the individual science exposures after masking out
all objects.  Frames were then flat-fielded and sky-subtracted.  The
images were de-warped to correct for known camera distortion. The
frames were aligned by centroiding on objects in the field; each frame
was then drizzled to a common output frame, and these output frames
were median-combined to produce the final image.

An empirical model for the PSF was derived from observations of
stars. In some cases a star was close enough to the lens galaxy to
fall in the NIRC2 field of view and be used as a model PSF. In other
cases a PSF star pair was used; this separate observation involves
observing a star cluster, and using as a PSF model a star that is at
the same distance from its tip-tilt correction star as the lens galaxy
was from its tip tilt star.  The star pair observations were made
immediately following the lens observations.


\section{Sample properties}
\label{sec:sampprop}

\begin{figure*}
\centerline{
\includegraphics[width=0.99\linewidth]{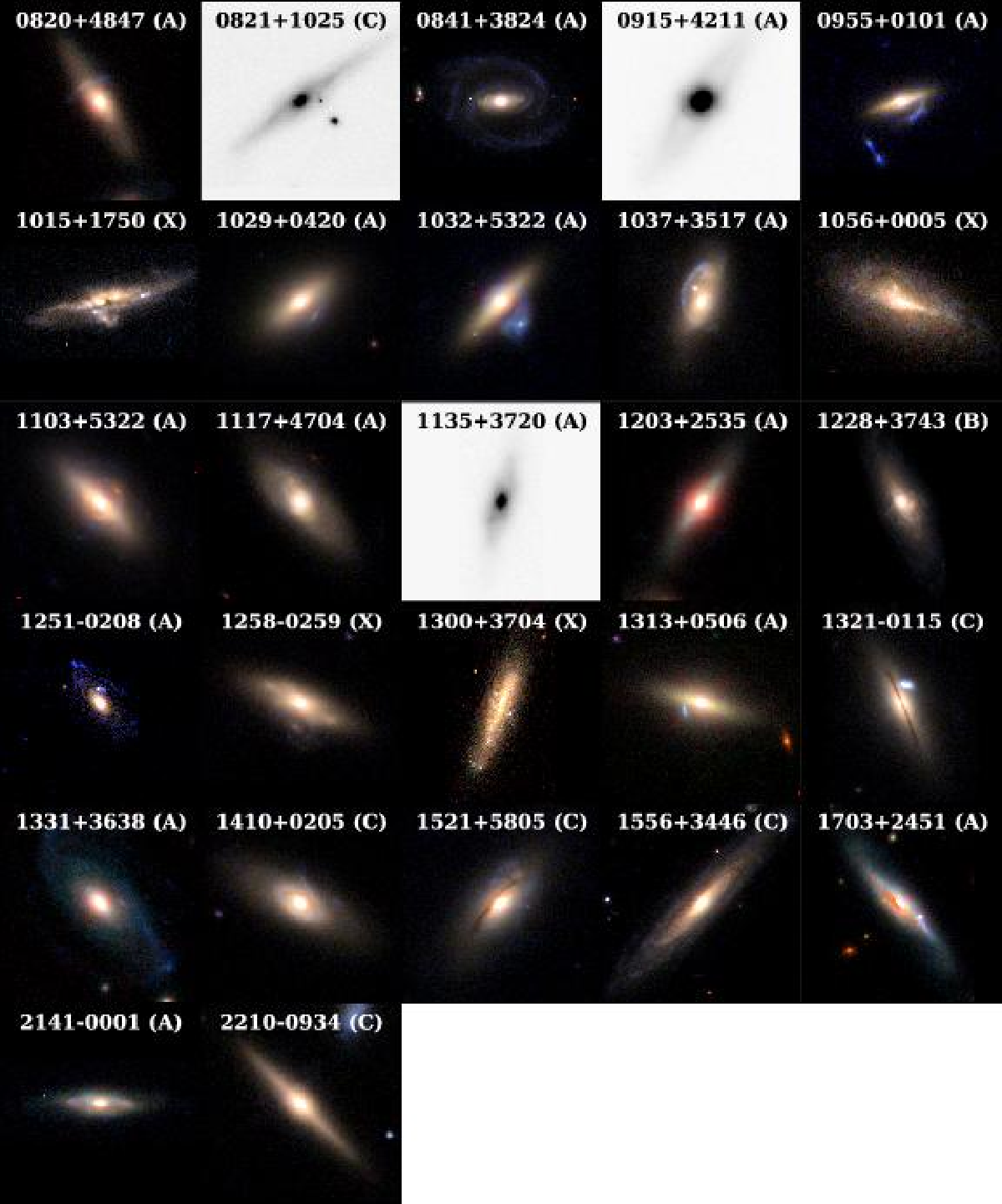}}
\caption{Colour montage of the high resolution images of
  all the SWELLS targets. The image sizes
  vary from $10^{\prime\prime}$ to $1^{\prime}$ on a side, and each lens'
  grade is shown in parentheses next to its name. 
  The three systems with grey-scale
  images only have AO K-band imaging. North is up, East is left. 
\label{fig:cmontage}}
\end{figure*}

In this section, we describe the overall properties of the
sample. First, in \Sref{ssec:lclass}, we present our classification of
the lensing morphology, providing notes on each of the systems based
on visual inspection of multi-colour images and deflector subtracted
images. These images are presented in Figures~\ref{fig:cmontage} and
\ref{fig:bspline1}. We note that subtracting the light of the 
deflector galaxy is much more challenging for the spiral galaxies of
the SWELLS sample, compared, e.g., to the smoother early-type galaxies
that compose the SLACS sample. For this reason the deflector
subtracted images shown in this paper are used only for the
classification of lensing morphology. In future papers, we will
discuss alternative strategies to disentangle light from the source
and the deflector for the purpose of lens modeling.

Second, in \Sref{ssec:morph} we describe the morphological and
structural properties of each deflector galaxy, based on visual
inspection and of multi-component, multi-band surface
photometry. Third, in~\Sref{ssec:mass}, we derive stellar mass for the
bulge and disk component of each of the deflectors.  We conclude in
\Sref{ssec:success} by investigating the dependency of the lens
confirmation rate on the properties of the deflector and the lensing
geometry.

\begin{figure*}
\centerline{
\includegraphics[width=0.79\linewidth]{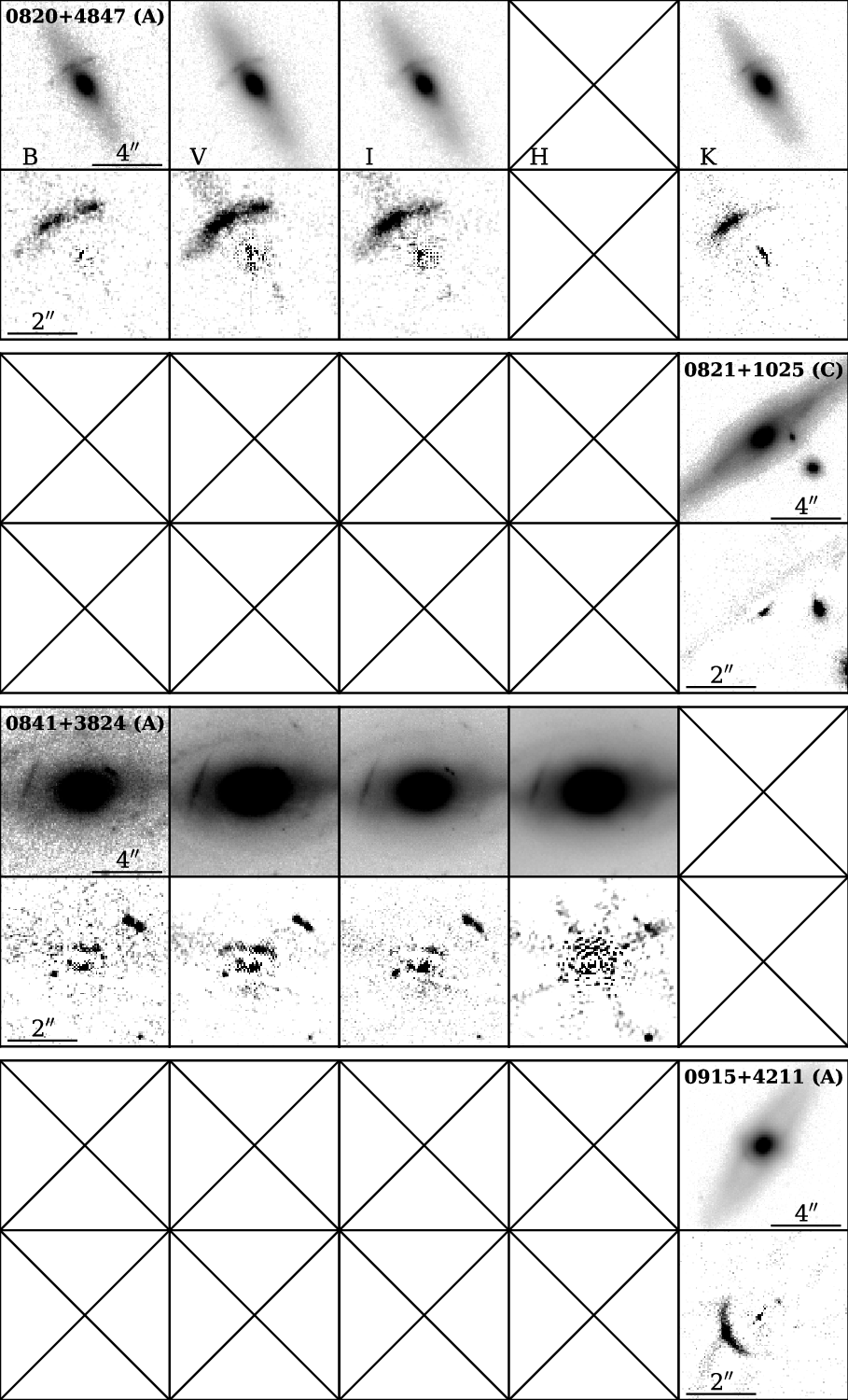}}
\caption{Single band high resolution 
images of the SWELLS targets. For each system the
top row shows the original image, and the bottom row shows the residual
image after B-spline subtraction of the deflector. Missing bands are
replaced by a cross. The top images are $10^{\prime\prime}$ on a side,
while the residual images are $5^{\prime\prime}$ on a side to show the
details of the potential sources identified in the SDSS spectroscopy. 
North is up, East is left.}
\label{fig:bspline1}
\end{figure*}

\begin{figure*}
\centerline{
\includegraphics[width=0.79\linewidth]{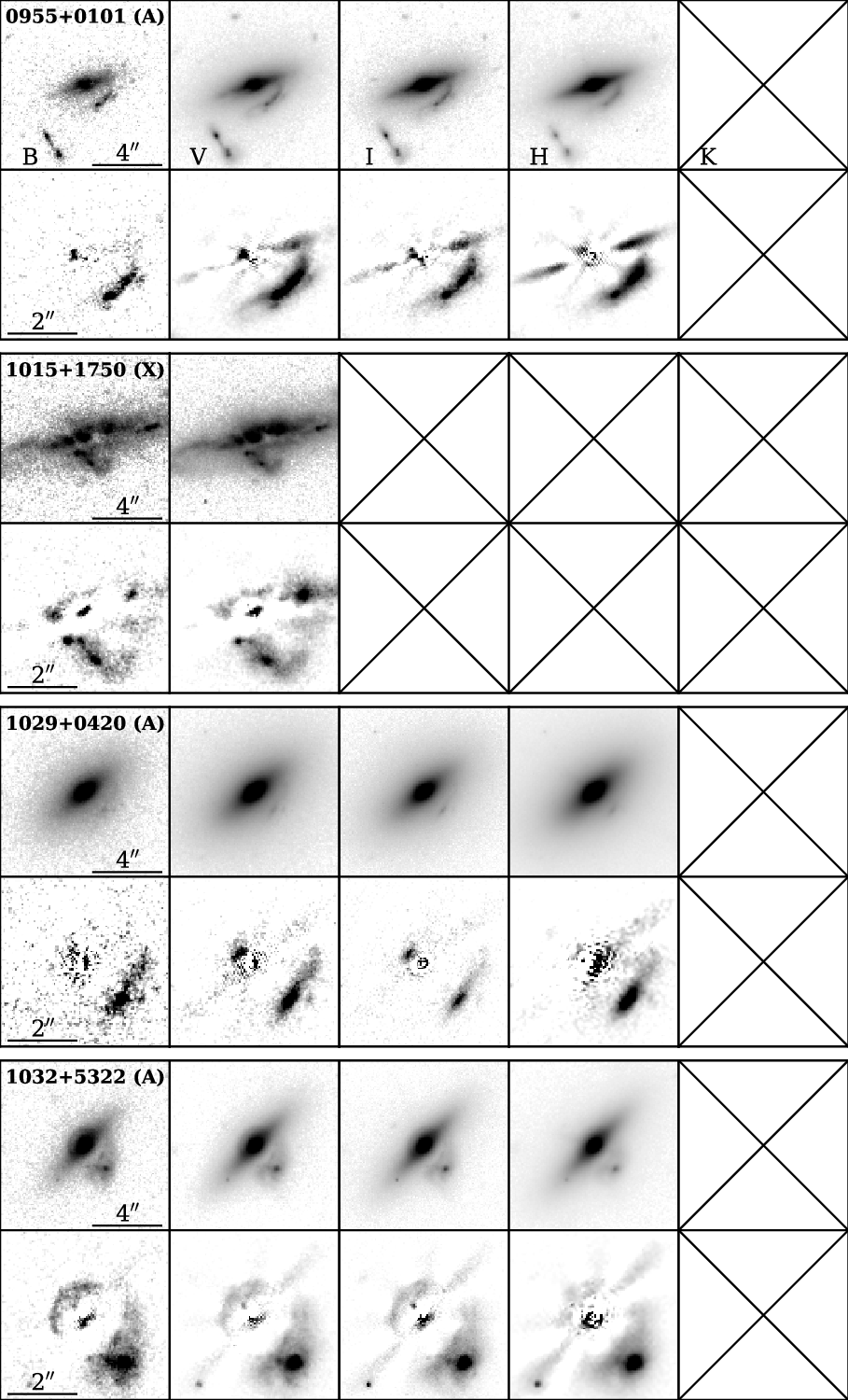}}
\contcaption{}
\end{figure*}

\begin{figure*}
\centerline{
\includegraphics[width=0.79\linewidth]{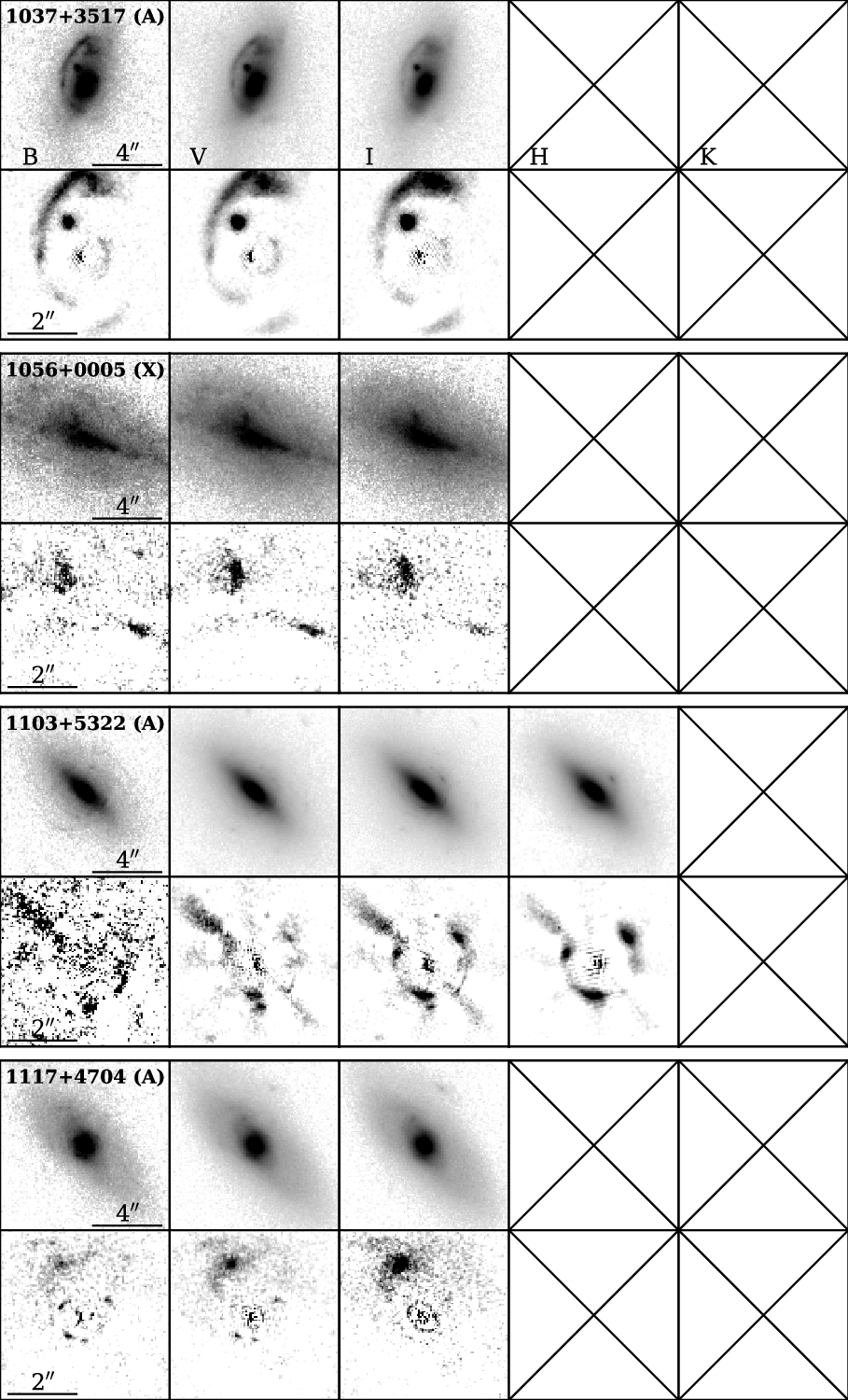}}
\contcaption{}
\end{figure*}

\begin{figure*}
\centerline{
\includegraphics[width=0.79\linewidth]{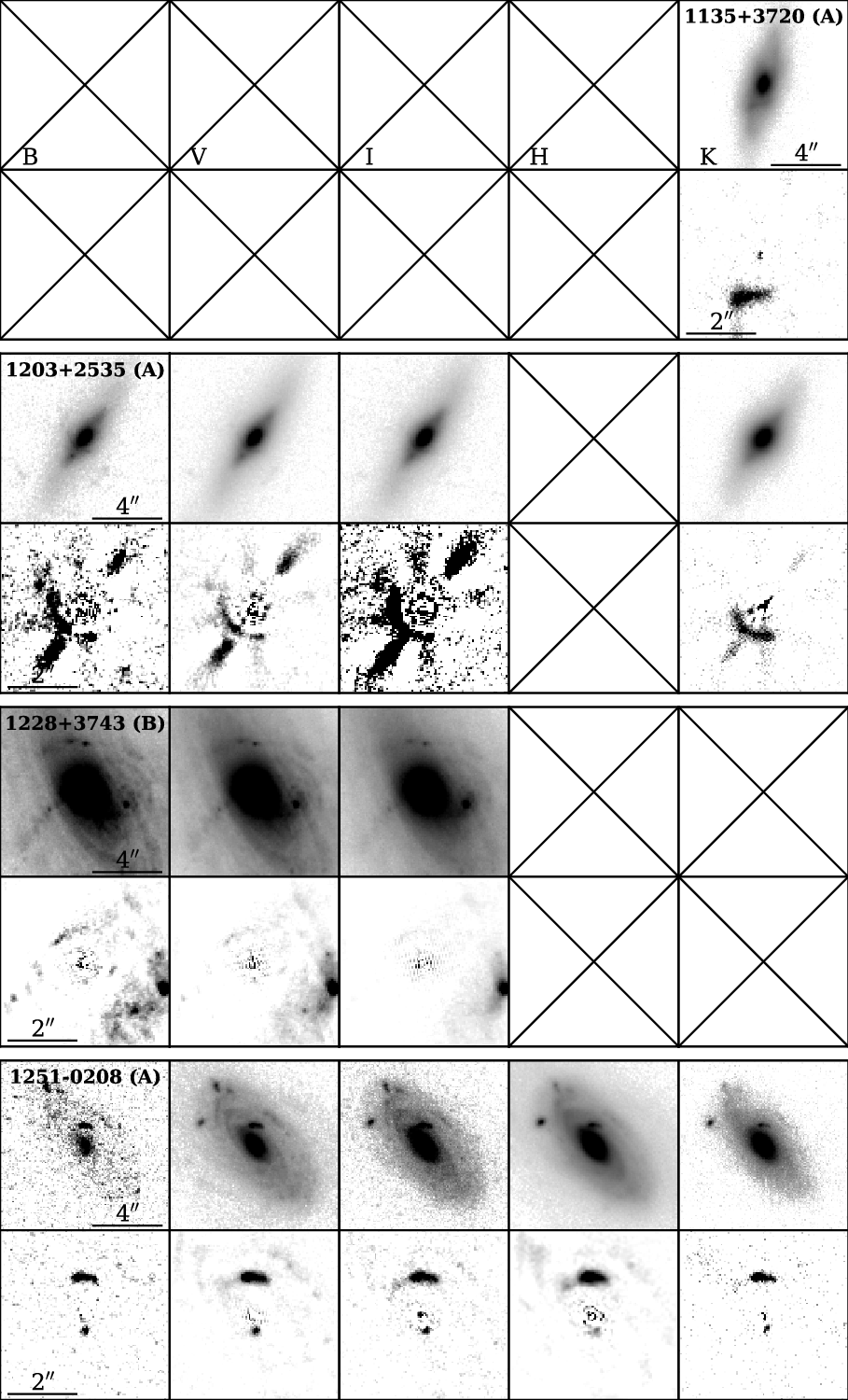}}
\contcaption{}
\end{figure*}

\begin{figure*}
\centerline{
\includegraphics[width=0.79\linewidth]{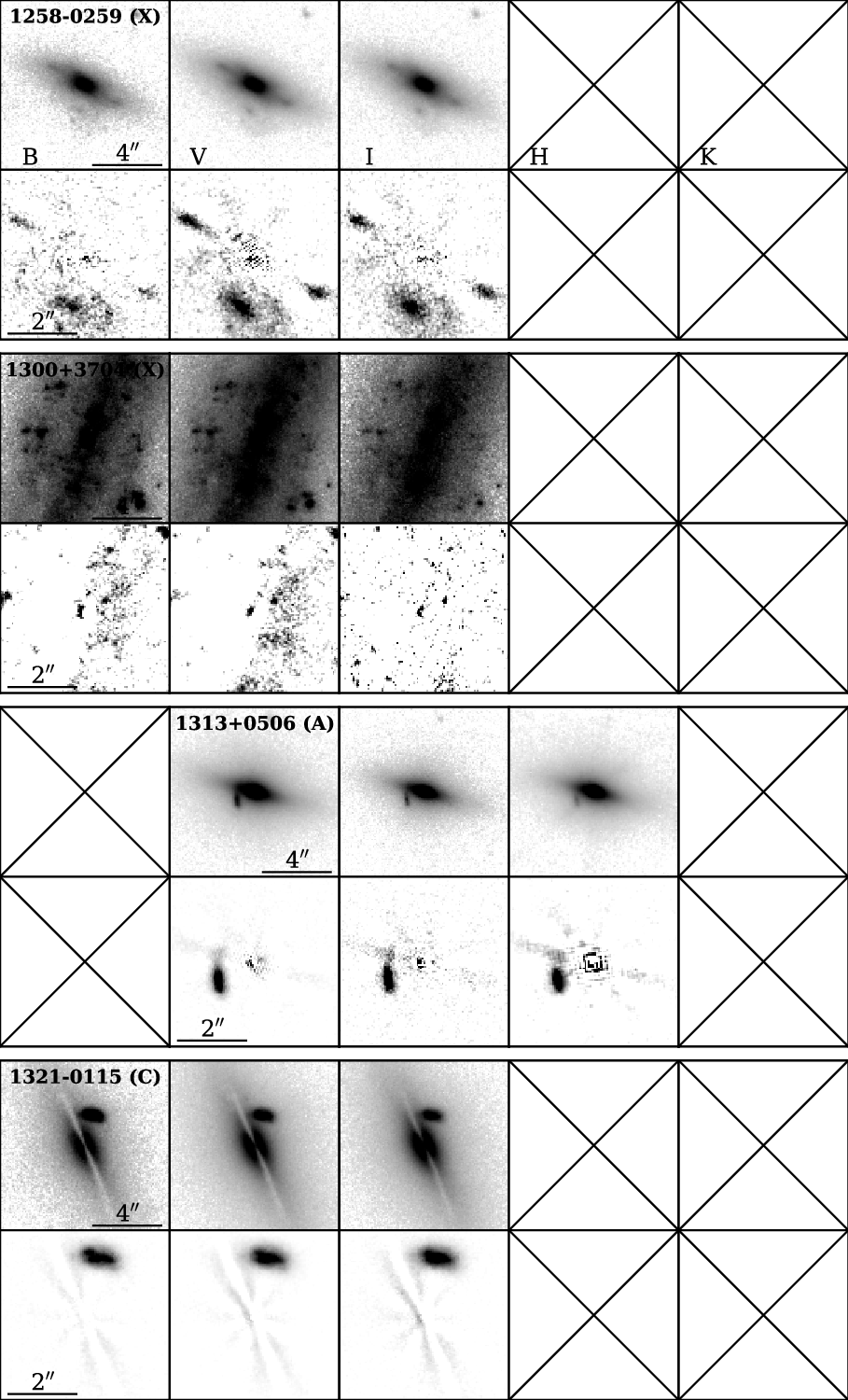}}
\contcaption{}
\end{figure*}

\begin{figure*}
\centerline{
\includegraphics[width=0.79\linewidth]{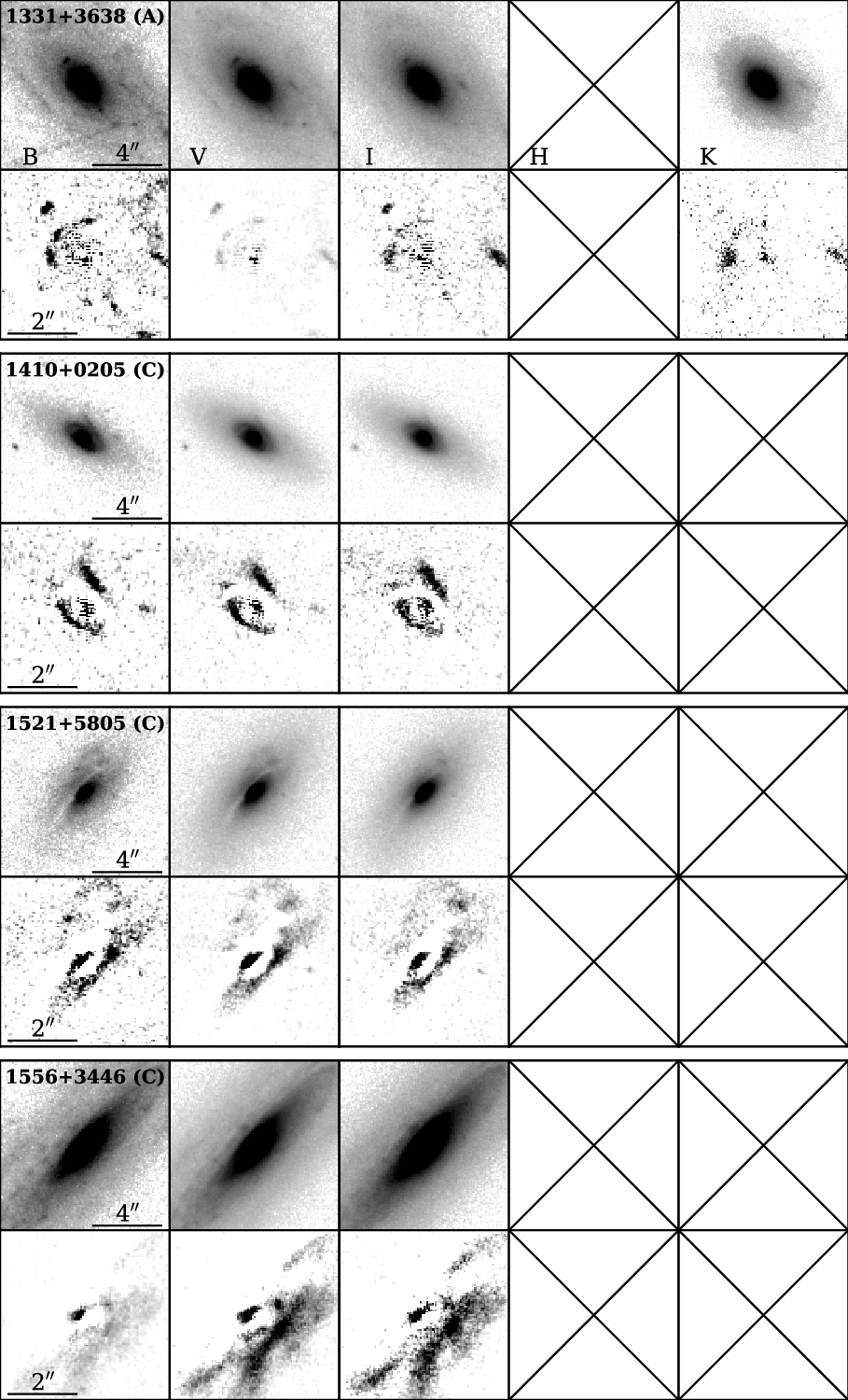}}
\contcaption{}
\end{figure*}

\begin{figure*}
\centerline{
\includegraphics[width=0.79\linewidth]{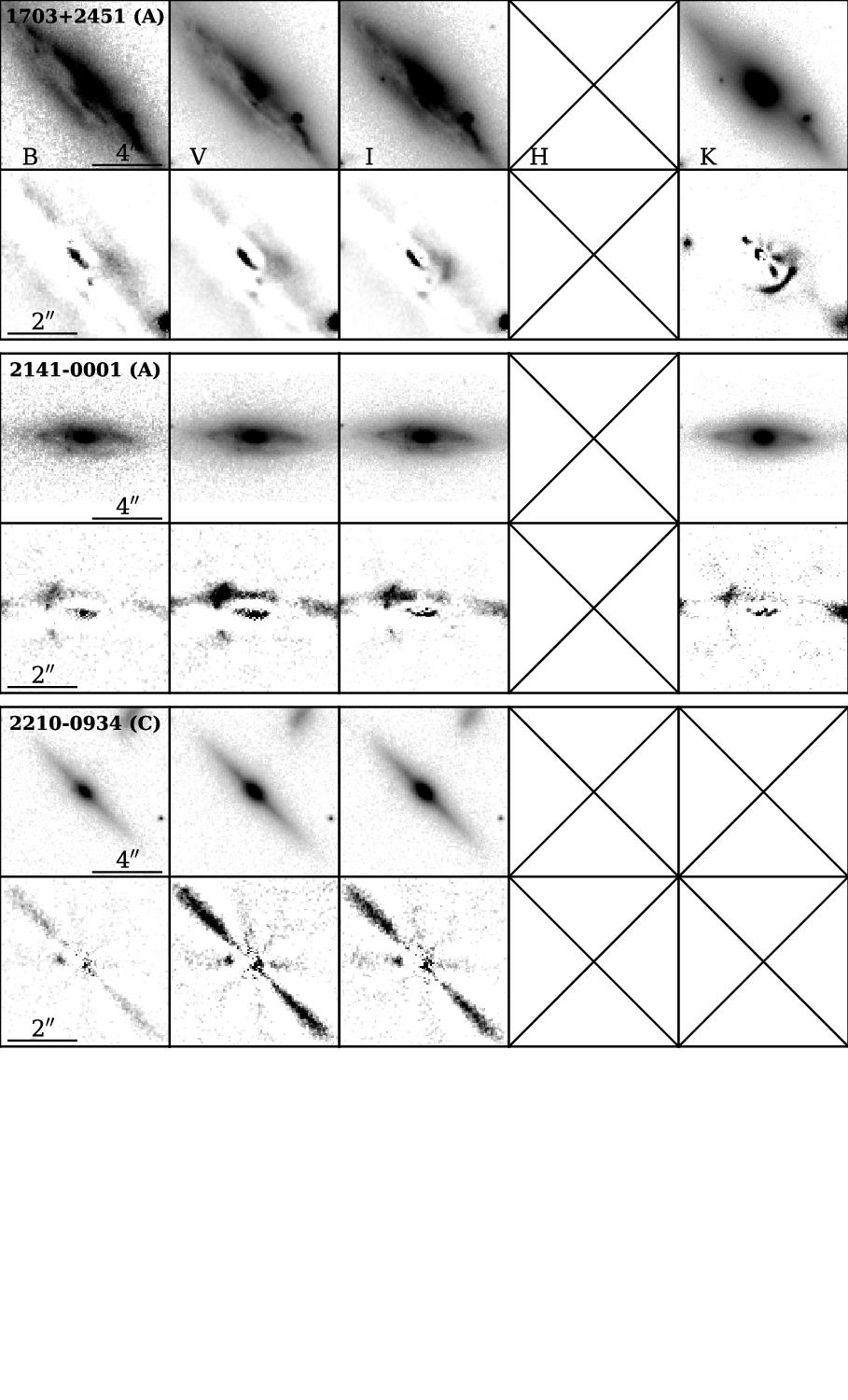}}
\contcaption{}
\end{figure*}

\subsection{Lensing classification}
\label{ssec:lclass}

A summary of the SWELLS sample is given in Table~\ref{tab:sample}.
All lenses were given a classification based on the scheme and criteria
defined by the SLACS survey \citep{Bol++06,Bol++08,Aug++09}. By
consensus of a subset of the authors (TT, AAD, MWA, PJM, BJB), lenses
were classified as secure (A-grade), probable (B-grade), possible
(C-grade) and not a lens (X-grade). We note that our classification
scheme is quite strict, and requires the convincing identification of
multiple images. It is likely that most of the B-grade lenses and some
of the C-grade systems could turn into A-grade systems if
deeper, or higher resolution, data were available to identify multiple
images.

In addition to the known late-type high-elongation lenses from the
SLACS survey (pre-Cycle~16s), \NSWELLSA\ A-grade previously unknown
strong gravitational lenses were discovered as part of this
project. Of the remaining eleven systems, there is \NSWELLSB\ B-grade
and \NSWELLSC\ C-grade systems. \NSWELLSX\ systems were classified as
definitely not-lenses. Full gravitational lens models for the secure
systems will be presented in future papers in this series (paper~II;
Brewer et al.\ 2011, in preparation).

\subsubsection{Notes on individual systems}
\label{sssec:notes}

{\bf 0820+4847} Grade A. Three component arc to NE of bulge. Deflector
has almost edge-on disk. Bulge is small, suggesting a later type
spiral, although spiral arms are not visible due to
inclination. Observed with LGSAO, source appears as two lensed images,
possibly indicating that the source has a red bulge and an extended
blue disk.

{\bf 0821+1025} Grade C. Observed only with LGSAO. Foreground galaxy
is high inclination disk. Background galaxy is compact, 1.7'' to the
W. The background galaxy might be strongly lensed at fainter surface
brightness levels. Blue band imaging might reveal multiple
images. There is also another source, possibly a satellite of the main
spiral galaxy, $\sim 3''$ SW of center.

{\bf 0841+3824} Grade A. The deflector is a grand design spiral with
blue arms and a redder bar/disk. A simple lens model, reproducing the
curved lensed feature, $\sim 1.5''$ NW of center and the counter image
on the opposite side, is given in the paper by \citet{Bol++08}.

{\bf 0915+4211} Grade A. Observed only with LGSAO. Clearly visible
arc and counter arc.

{\bf 0955+0101} Grade A. The deflector is an edge-on disky galaxy. A
blue and highly sheared arc is visible to the SW of the bulge and the
counter-image is clearly visible in the residuals, especially the
B-band. A simple lens model is given in the paper by
\citet{Bol++08}. Another source, possibly a satellite of the
deflector, or of the background source, is visible $\sim 5''$ to the S
of the deflector.

{\bf 1015+1750} Grade X. The deflector is a dusty star forming edge-on
spiral. The likely background galaxy is $1.6''$ to SE. No F814W
image. No evidence for highly sheared images.

{\bf 1029+0420} Grade A. The deflector is smooth and disky and seen
edge-on, likely an early-type spiral or lenticular galaxy for the
relatively large bulge. Arc and counter arc are clearly visible in the
colour image and in the bspline subtracted image.  A simple lens model is
given in the paper by \citet{Bol++08}.

{\bf 1032+5322} Grade A. The deflector is smooth and disky and seen
edge-on. The lensed source is blue and has a complex morphology,
likely a spiral with a redder bulge. Images and counter-images are
clearly visible. A simple lens model is given in the paper by
\citet{Bol++08}.

{\bf 1037+3517} Grade A. The main deflector is a spiral
galaxy. However, a second peak in red surface brightness indicates
either a merger in process or a multi-galaxy lens, if the two galaxies
are not in the same plane. Arc and counter arc are blue and clearly
visible, although the morphology of the lensed images is not
straightforward to interpret, owing to the complex lens
potential. Could be a multi-galaxy lens, possible 2nd (compact red)
lens galaxy at 1.1'' to N. Extended source galaxy 2.4'' to N.
Integral field spectroscopy or high resolution narrow band imaging
would help decipher this configuration.

{\bf 1056+0005} Grade X. Deflector galaxy is a bulge-less star forming
spiral. Likely background galaxy is seen $1\farcs3$ North of
centre. No evidence of shear suggests that the source is well outside
the caustics of the deflector.

{\bf 1103+5322} Grade A. The deflector is smooth and disky and seen
edge-on. The lensed source is blue and has a complex morphology,
likely a spiral with a redder bulge. Images and counter-images are
clearly visible. A simple lens model is given in the paper by
\citet{Bol++08}.

{\bf 1117+4704} Grade A. The deflector is smooth and disky and seen
close to edge-on. A bright and highly sheared source is visible
1$\farcs6$ NE of deflector centre, albeit with no clear evidence for a
counter-image. A classic ``fold'' quadruply imaged source is visible
in the B-band and V-band b-spline residuals, with radius approximately
$0\farcs75$.

{\bf 1135+3720} Grade A. Observed only with LGSAO. Arc and
counter-image are clearly visible in the b-spline residuals.

{\bf 1203+2535} Grade A. Deflector galaxy has a blue edge-on disk and
a red bulge; the outer edge of the disk is warped. Arc made of three
merging images is clearly visible SE of bulge.  Possible quad with
faint counter-image near the centre or naked cusp configuration.

{\bf 1228+3743} Grade B. Deflector is a grand design
spiral. Background source is also extended with a red bulge and blue
spiral arms, one of which passes close to the lens bulge. A possible
arc and counter arc with image separation $\sim 2''$ is visible in the
residuals, clearest in the B-band.

{\bf 1251-0208} Grade A. The deflector has a red bulge and blue
spirals arms. The lensed source is blue and high surface
brightness. Images and counter-images are clearly visible in the
b-spline residuals. Simple lens models are given in the papers by
\citet{Bol++06,Bol++08}. A lensing and dynamics analysis is presented
in the paper by \citet{Bar++11}.

{\bf 1258-0259} Grade X. Foreground galaxy has edge-on disk and bar
like structure. A possible background galaxy is seen $1\farcs6$ S of
the deflector, with bulge and spiral arms. The source does not appear
to be sheared and no counter-image candidate could be identified.

{\bf 1300+3704} Grade X. Foreground galaxy is a large bulge-less
edge-on star forming spiral. Possible background galaxies $3\farcs5$
to NE and SW of deflector. 

{\bf 1313+0506} Grade A. The deflector is smooth and disky and seen
edge-on. The lensed source is blue and high surface brightness.
Images and counter-images are clearly visible in the residuals. A simple lens
model is given in the paper by \citet{Bol++08}.

{\bf 1321-0115} Grade C. Foreground galaxy has edge-on disk with
prominent dust lane. Blue background galaxy is seen $1\farcs9$ to N of
the deflector. The source is elongated tangentially to the direction
of the deflector suggesting strong shear. No counter-image could be
identified, although it could be obscured by the prominent dust
lane. High resolution infrared images would be useful to confirm or
reject the lensing hypothesis.

{\bf 1331+3638} Grade A. Deflector galaxy has blue spiral arms and a
red bulge. Arc made of three merging images is clearly visible NE of
bulge.  Possible quad with faint counter-image near the centre or naked
cusp configuration.  The target was observed with LGSAO, but faint
tip-tilt star yielded poor correction.

{\bf 1410+0205} Grade C. The deflector has a red bulge, a prominent
disk and a ring-like blue structure surrounding the bulge. The blue
structure is most likely due to star formation in the deflector
although lensing cannot be ruled out with current data. Another
possible source is seen NW of lens centre, although direction of the
elongation is not obviously consistent with strong lensing. It could
be that the source is only partially multiply imaged with the bulk of
it extending outside of the caustic. Integral field spectroscopy or
narrow band imaging would be useful to decipher this system.

{\bf 1521+5805} Grade C. Foreground galaxy has nearly edge-on disk
with prominent dust lane. The background galaxy, with bulge and spiral
arms, is visible NW of deflector. It is possible that parts of the
background source are multiply imaged but the dust lane and the
complex morphology prevent a conclusive answer. High resolution
infrared images and integral field spectroscopy or narrow band
imaging would help. 

{\bf 1556+3446} Grade C. The deflector is a dusty nearly edge-on
spiral galaxy. A possible image and counter-image is visible in the
b-spline residuals. In the SDSS spectrum, background source has clear OII,
Hb, OIII emission lines.  High resolution K-band imaging would help
see through the dust, while narrow band imaging or integral field
spectroscopy would help disentangle the light from the source and from
the deflector.

{\bf 1703+2451} Grade A.  The deflector is an edge-on dusty spiral,
with boxy isophotes suggesting a pseudo bulge. The outer disk is
warped. A red(dened) arc and possible counter-image are clearly
visible, especially in the LGSAO K-band b-spline residuals.

{\bf 2141-0001} Grade A. Described in detail in paper II
\citep{Dut++11b}, and more simply by \citet{Bol++08}.

{\bf 2210-0934} Grade C.  The deflector has a prominent edge-on disk.
Possible image and counter-images are visible in the b-spline
residuals, with image separation $\sim1\farcs3$. Another source,
possibly a satellite of the main lens, is visible $\sim5''$ N of
center.

\begin{table*}
\begin{center}
\scriptsize\begin{tabular}{lcccccccccccc} 
\hline
\hline 
{\bf ID} & B$_{\rm WFPC2}$ &    B$_{\rm ACS}$  & V$_{\rm WFPC2}$  & I$_{\rm WFPC2}$  &  I$_{\rm ACS}$ &  H$_{\rm NIC2}$ &  H$_{\rm WFC3}$  &  K$'_{\rm AO}$  & R$_{\rm e}$/kpc  &   $q$   &  $\log$(M$_{*,\rm Chab}$/M$_\odot$) &    $\log$(M$_{*,\rm Salp}$/M$_\odot$)\\
\hline
J0820+4847 &   19.84 & \nodata &   18.62 &   17.95 & \nodata & \nodata & \nodata &   16.85 &  0.73 & 0.58 & 10.58$\pm$0.09 & 10.80$\pm$0.09 \\
           &   19.67 & \nodata &   18.44 &   17.86 & \nodata & \nodata & \nodata &   17.20 &  3.05 & 0.25 & 10.45$\pm$0.09 & 10.70$\pm$0.08 \\
J0821+1025 & \nodata & \nodata & \nodata & \nodata & \nodata & \nodata & \nodata &   16.12 &  1.56 & 0.58 & 10.63$\pm$0.14 & 10.90$\pm$0.13 \\
           & \nodata & \nodata & \nodata & \nodata & \nodata & \nodata & \nodata &   15.64 &  4.07 & 0.12 & 10.84$\pm$0.13 & 11.10$\pm$0.15 \\
J0841+3824 &   18.29 & \nodata &   17.08 & \nodata &   16.36 & \nodata &   15.46 & \nodata &  2.42 & 1.00 & 11.05$\pm$0.10 & 11.32$\pm$0.09 \\
           &   17.43 & \nodata &   16.25 & \nodata &   15.66 & \nodata &   15.05 & \nodata & 17.53 & 0.56 & 11.23$\pm$0.10 & 11.46$\pm$0.09 \\
J0915+4211 & \nodata & \nodata & \nodata & \nodata & \nodata & \nodata & \nodata &   16.46 &  1.06 & 0.66 & 10.38$\pm$0.12 & 10.61$\pm$0.11 \\
           & \nodata & \nodata & \nodata & \nodata & \nodata & \nodata & \nodata &   16.31 &  1.06 & 0.74 & 10.41$\pm$0.14 & 10.68$\pm$0.13 \\
J0955+0101 & \nodata &   20.43 &   18.20 & \nodata &   17.47 &   16.53 & \nodata & \nodata &  1.62 & 0.57 & 10.63$\pm$0.09 & 10.85$\pm$0.09 \\
           & \nodata &   20.45 &   19.09 & \nodata &   18.25 &   17.43 & \nodata & \nodata &  1.62 & 0.19 & 10.17$\pm$0.08 & 10.44$\pm$0.09 \\
J1015+1750 & \nodata & \nodata & \nodata & \nodata & \nodata & \nodata & \nodata & \nodata & \nodata & \nodata & \nodata &  \nodata \\
           &   18.88 & \nodata &   17.73 & \nodata & \nodata & \nodata & \nodata & \nodata &  4.95 & 0.26 & 10.70$\pm$0.22 & 10.95$\pm$0.20 \\
J1029+0420 &   18.18 & \nodata &   17.06 & \nodata &   16.39 & \nodata &   15.56 & \nodata &  2.13 & 0.48 & 10.93$\pm$0.09 & 11.17$\pm$0.10 \\
           &   19.49 & \nodata &   18.42 & \nodata &   17.87 & \nodata &   17.47 & \nodata &  4.20 & 0.82 & 10.18$\pm$0.10 & 10.44$\pm$0.09 \\
J1032+5322 &   19.59 & \nodata &   18.31 & \nodata &   17.58 & \nodata &   16.63 & \nodata &  1.21 & 0.66 & 10.73$\pm$0.09 & 10.96$\pm$0.08 \\
           &   20.13 & \nodata &   18.90 & \nodata &   18.24 & \nodata &   17.47 & \nodata &  1.97 & 0.22 & 10.39$\pm$0.09 & 10.63$\pm$0.10 \\
J1037+3517 &   18.98 & \nodata &   17.72 &   17.06 & \nodata & \nodata & \nodata & \nodata &  2.78 & 0.68 & 10.76$\pm$0.14 & 11.03$\pm$0.13 \\
           &   19.48 & \nodata &   18.43 &   17.88 & \nodata & \nodata & \nodata & \nodata &  3.11 & 0.36 & 10.30$\pm$0.13 & 10.58$\pm$0.12 \\
J1056+0005 & \nodata & \nodata & \nodata & \nodata & \nodata & \nodata & \nodata & \nodata &  \nodata & \nodata & \nodata &  \nodata \\
           &   18.18 & \nodata &   17.41 &   17.03 & \nodata & \nodata & \nodata & \nodata &  2.43 & 0.47 &  9.60$\pm$0.10 &  9.82$\pm$0.13 \\
J1103+5322 &   19.18 & \nodata &   17.79 & \nodata &   17.12 &   16.32 & \nodata & \nodata &  3.62 & 0.56 & 10.99$\pm$0.09 & 11.23$\pm$0.09 \\
           &   19.42 & \nodata &   18.37 & \nodata &   17.65 &   16.92 & \nodata & \nodata &  3.62 & 0.37 & 10.72$\pm$0.09 & 10.95$\pm$0.10 \\
J1117+4704 &   19.85 & \nodata &   18.56 &   17.90 & \nodata & \nodata & \nodata & \nodata &  1.24 & 1.00 & 10.69$\pm$0.14 & 10.94$\pm$0.16 \\
           &   18.89 & \nodata &   17.69 &   17.14 & \nodata & \nodata & \nodata & \nodata &  5.24 & 0.38 & 10.93$\pm$0.15 & 11.17$\pm$0.12 \\
J1135+3720 & \nodata & \nodata & \nodata & \nodata & \nodata & \nodata & \nodata &   17.70 &  2.96 & 1.00 & 10.42$\pm$0.13 & 10.68$\pm$0.13 \\
           & \nodata & \nodata & \nodata & \nodata & \nodata & \nodata & \nodata &   16.37 &  2.98 & 0.38 & 10.97$\pm$0.13 & 11.23$\pm$0.14 \\
J1203+2535 &   19.08 & \nodata &   17.99 &   17.34 & \nodata & \nodata & \nodata &   16.77 &  0.57 & 0.53 & 10.43$\pm$0.10 & 10.67$\pm$0.08 \\
           &   19.00 & \nodata &   18.08 &   17.60 & \nodata & \nodata & \nodata &   16.70 &  2.53 & 0.32 & 10.39$\pm$0.08 & 10.61$\pm$0.09 \\
J1228+3743 &   17.18 & \nodata &   16.16 &   15.50 & \nodata & \nodata & \nodata & \nodata &  1.10 & 0.97 & 10.37$\pm$0.12 & 10.62$\pm$0.13 \\
           &   16.05 & \nodata &   15.26 &   14.77 & \nodata & \nodata & \nodata & \nodata &  3.99 & 0.31 & 10.56$\pm$0.13 & 10.78$\pm$0.11 \\
J1251$-$0208 & \nodata &   21.92 &   19.90 & \nodata &   18.87 & \nodata &   17.81 &   17.57 &  1.68 & 0.50 & 10.68$\pm$0.09 & 10.95$\pm$0.07 \\
           & \nodata &   20.10 &   18.36 & \nodata &   17.80 & \nodata &   17.02 &   16.98 &  8.49 & 0.50 & 10.96$\pm$0.08 & 11.21$\pm$0.08 \\
J1258$-$0259 &   19.73 & \nodata &   19.81 &   17.97 & \nodata & \nodata & \nodata & \nodata &  1.56 & 0.62 & 10.10$\pm$0.14 & 10.29$\pm$0.14 \\
           &   19.33 & \nodata &   19.15 &   17.70 & \nodata & \nodata & \nodata & \nodata &  2.55 & 0.32 & 10.19$\pm$0.13 & 10.41$\pm$0.13 \\
J1300+3704 & \nodata & \nodata & \nodata & \nodata & \nodata & \nodata & \nodata & \nodata &  \nodata & \nodata & \nodata &  \nodata \\
           &   16.29 & \nodata &   15.81 &   15.53 & \nodata & \nodata & \nodata & \nodata &  2.21 & 0.27 &  9.38$\pm$0.11 &  9.64$\pm$0.10 \\
J1313+0506 & \nodata & \nodata &   17.95 & \nodata &   17.26 &   16.50 & \nodata & \nodata &  1.64 & 0.40 & 10.83$\pm$0.09 & 11.05$\pm$0.10 \\
           & \nodata & \nodata &   19.29 & \nodata &   18.89 &   17.99 & \nodata & \nodata &  5.89 & 0.99 & 10.17$\pm$0.13 & 10.41$\pm$0.11 \\
J1321$-$0115 &   18.99 & \nodata &   17.73 &   17.00 & \nodata & \nodata & \nodata & \nodata &  4.88 & 1.00 & 10.68$\pm$0.13 & 10.96$\pm$0.13  \\
           &   18.36 & \nodata &   17.31 &   16.76 & \nodata & \nodata & \nodata & \nodata &  4.90 & 0.35 & 10.68$\pm$0.16 & 10.93$\pm$0.12 \\
J1331+3638 &   18.62 & \nodata &   17.42 &   16.76 & \nodata & \nodata & \nodata &   15.70 &  2.86 & 0.66 & 10.89$\pm$0.09 & 11.16$\pm$0.10 \\
           &   18.23 & \nodata &   17.42 &   17.06 & \nodata & \nodata & \nodata &   17.15 &  7.88 & 0.40 & 10.46$\pm$0.10 & 10.72$\pm$0.08 \\
J1410+0205 &   19.99 & \nodata &   18.76 &   18.08 & \nodata & \nodata & \nodata & \nodata &  0.98 & 0.68 & 10.41$\pm$0.14 & 10.62$\pm$0.15 \\
           &   19.59 & \nodata &   18.55 &   18.02 & \nodata & \nodata & \nodata & \nodata &  2.66 & 0.37 & 10.33$\pm$0.15 & 10.55$\pm$0.14 \\
J1521+5805 &   19.89 & \nodata &   18.37 &   17.49 & \nodata & \nodata & \nodata & \nodata &  3.55 & 0.59 & 11.10$\pm$0.12 & 11.37$\pm$0.13 \\
           &   19.89 & \nodata &   18.60 &   18.20 & \nodata & \nodata & \nodata & \nodata &  7.40 & 0.46 & 10.67$\pm$0.12 & 10.93$\pm$0.12 \\
J1556+3446 &   17.98 & \nodata &   16.85 &   16.16 & \nodata & \nodata & \nodata & \nodata &  2.04 & 0.45 & 10.66$\pm$0.16 & 10.89$\pm$0.14 \\
           &   17.30 & \nodata &   16.40 &   15.86 & \nodata & \nodata & \nodata & \nodata &  6.01 & 0.18 & 10.69$\pm$0.13 & 10.91$\pm$0.13 \\
J1703+2451 &   20.16 & \nodata &   18.55 &   17.46 & \nodata & \nodata & \nodata &   15.35 &  1.46 & 0.53 & 10.30$\pm$0.07 & 10.58$\pm$0.06 \\
           &   17.30 & \nodata &   16.33 &   15.74 & \nodata & \nodata & \nodata &   15.06 &  3.55 & 0.25 & 10.68$\pm$0.09 & 10.93$\pm$0.09 \\
J2141$-$0001 &   20.76 & \nodata &   19.28 & \nodata &   18.33 & \nodata & \nodata &   16.88 &  1.01 & 0.55 & 10.57$\pm$0.10 & 10.80$\pm$0.10 \\
           &   19.23 & \nodata &   17.98 & \nodata &   17.31 & \nodata & \nodata &   16.60 &  3.64 & 0.26 & 10.73$\pm$0.09 & 10.97$\pm$0.08 \\
J2210$-$0934 &   19.50 & \nodata &   18.50 &   17.93 & \nodata & \nodata & \nodata & \nodata &  0.81 & 0.62 & 10.04$\pm$0.15 & 10.22$\pm$0.14 \\
           &   19.60 & \nodata &   18.66 &   18.16 & \nodata & \nodata & \nodata & \nodata &  1.28 & 0.15 &  9.89$\pm$0.13 & 10.08$\pm$0.13 \\
\hline
\hline
\end{tabular}
\caption{Photometric, structural and stellar mass parameters of the
SWELLS systems.  For each lens the first line lists the bulge
parameters (when detected), while the second line lists the disk
parameters. Typical uncertainties are 0.05 mags on magnitudes, 10\% on
effective radii and axis ratios ($q=b/a$), dominated by systematics
\citep[see, e.g.,][and references therein]{Aug++09}. Apparent
magnitudes are not corrected for galactic extinction. Effective radii
are given along the intermediate axis; major-axis effective radii can
be obtained by dividing by $\sqrt{q}$.
\label{tab:structure}}
\end{center}
\end{table*}

\subsection{Morphological and structural properties of the deflectors}
\label{ssec:morph}

\begin{figure}
\centering\includegraphics[width=0.9\linewidth]{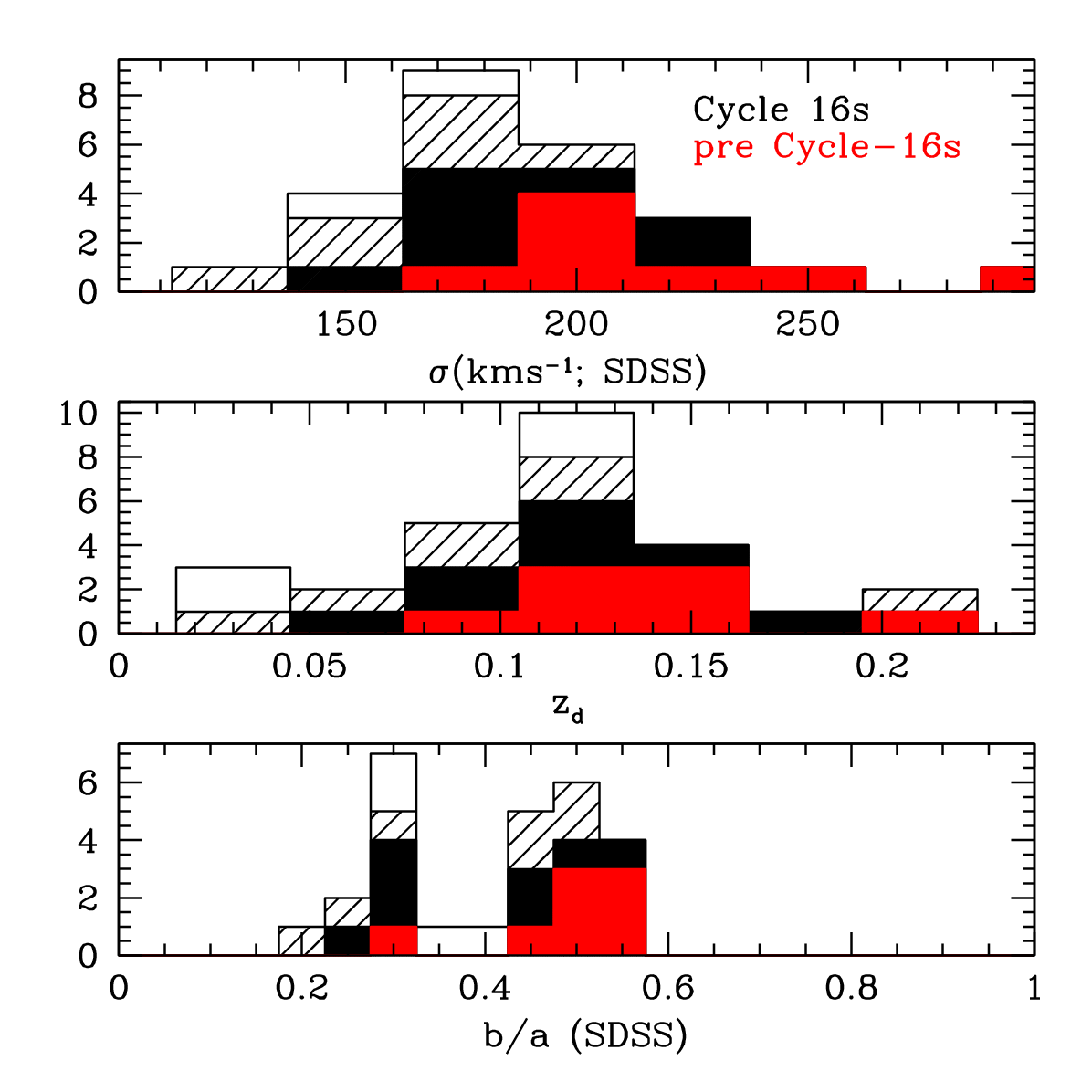}
\caption{Histogram of SDSS parameters for SWELLS lenses. The top panel
shows the distribution of stellar velocity dispersions (when available,
see Table~\ref{tab:sample}), 
the middle panel shows the distribution of deflector
redshifts, and the bottom panel shows the distribution of axis ratios.
Systems discovered prior to Cycle~16s are plotted in red, our new
systems are shown in black. Solid histograms represent secure lenses
(A), hatched histograms represent probable and possible lenses (B
and C), while open histograms represent non-lenses (X). Note that
our Cycle~16s strategy targeted preferentially lower velocity
dispersion and axis ratio systems, in order to select disk edge-on
galaxies. Note also how the confirmation rate increases with $\sigma$,
as expected. 
\label{fig:hist}}
\end{figure}

All the SWELLS deflectors show a clear disky component
(Figure~\ref{fig:cmontage}), confirming our original selection based
on SDSS images.  Furthermore, the targets have high projected
ellipticity, again consistent with our original selection based on
SDSS images (Figure~\ref{fig:hist}). Many of the targets show clear
spiral arms, which in several cases are blue. Other targets are so
close to edge-on that it is hard to identify spiral arms conclusively
(notes on the morphology of individual targets are provided in the
previous section).  The overall morphology of the targets is also
consistent with the distribution of velocity dispersions
(Figure~\ref{fig:hist}); this distribution has a lower central value
than the SLACS sample, which is dominated by massive elliptical
galaxies.

We quantify the structural properties of the deflectors by performing
a bulge-disk decomposition. For the purpose of this paper we adopt a
simple parametrisation, and describe the bulge with an elliptical
\citet{dev48} profile and the disk with an elliptical exponential
profile. This parametrisation is the same as that adopted by Simard et
al.\ (2011, in preparation) to analyse the entire SDSS-DR7 database,
and therefore is ideal for a comparison between lenses and non-lens
galaxies. Beyond this first analysis, in future papers we will derive
more accurate descriptions of the surface brightness distribution by
considering more flexible models, like, e.g., the Sersic profile for
the bulge (e.g. paper II), and by correcting for extinction due to
prominent dust lanes (Brewer et al. 2011, in preparation).

For each deflector we compute the posterior probability distribution
function of the parameters by comparing the models with the
multi-colour images using the dedicated code \SPASMOID developed by
one of us \citep[MWA; see also][]{Ben++11}. \SPASMOID replaces the
functionality of standard codes like \galfit, by enabling a full
exploration of the likelihood surface as well as joint analysis of
multiband data. Furthermore, SPASMOID models fit the multi-band
imaging simultaneously with the same effective radius, and the bulge
effective radius is constrained to be smaller than the disk effective
radius\footnote{The priors for the surface brightness models are
uniform but have meaningful bounds: $\pm$5 pixels from the brightest
central pixel for the centroid, 0.1 (0.15) to 1 for the axis ratio of
the disk (bulge), and effective radii between 0.1 arcsec (2 pixels)
and the size of the cutout that is fitted by SPASMOID (typically
hundreds of pixels). Furthermore, the effective radii, flattening, and
position angle are the same for each filter, while the centroids of
the bulge and disk are allowed a small shift of 3 pixels with respect
to each other.}. This procedure ensures self-consistent colors even
with data of varying depth as a function of wavelength. The most
probable values of the parameters are given in
Table~\ref{tab:structure}.

\begin{figure}
\centering\includegraphics[width=0.8\linewidth]{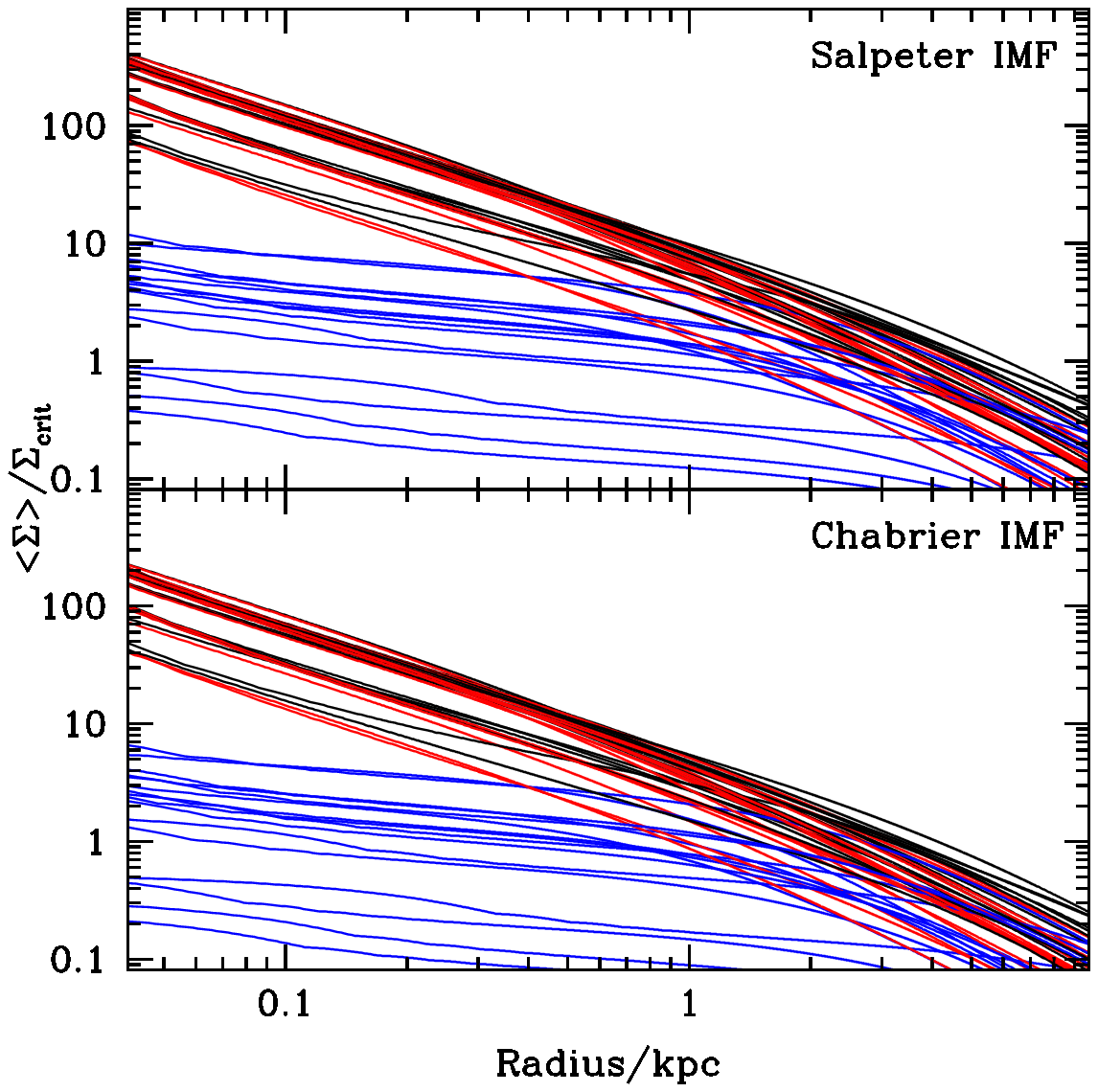}
\centering\includegraphics[width=0.8\linewidth]{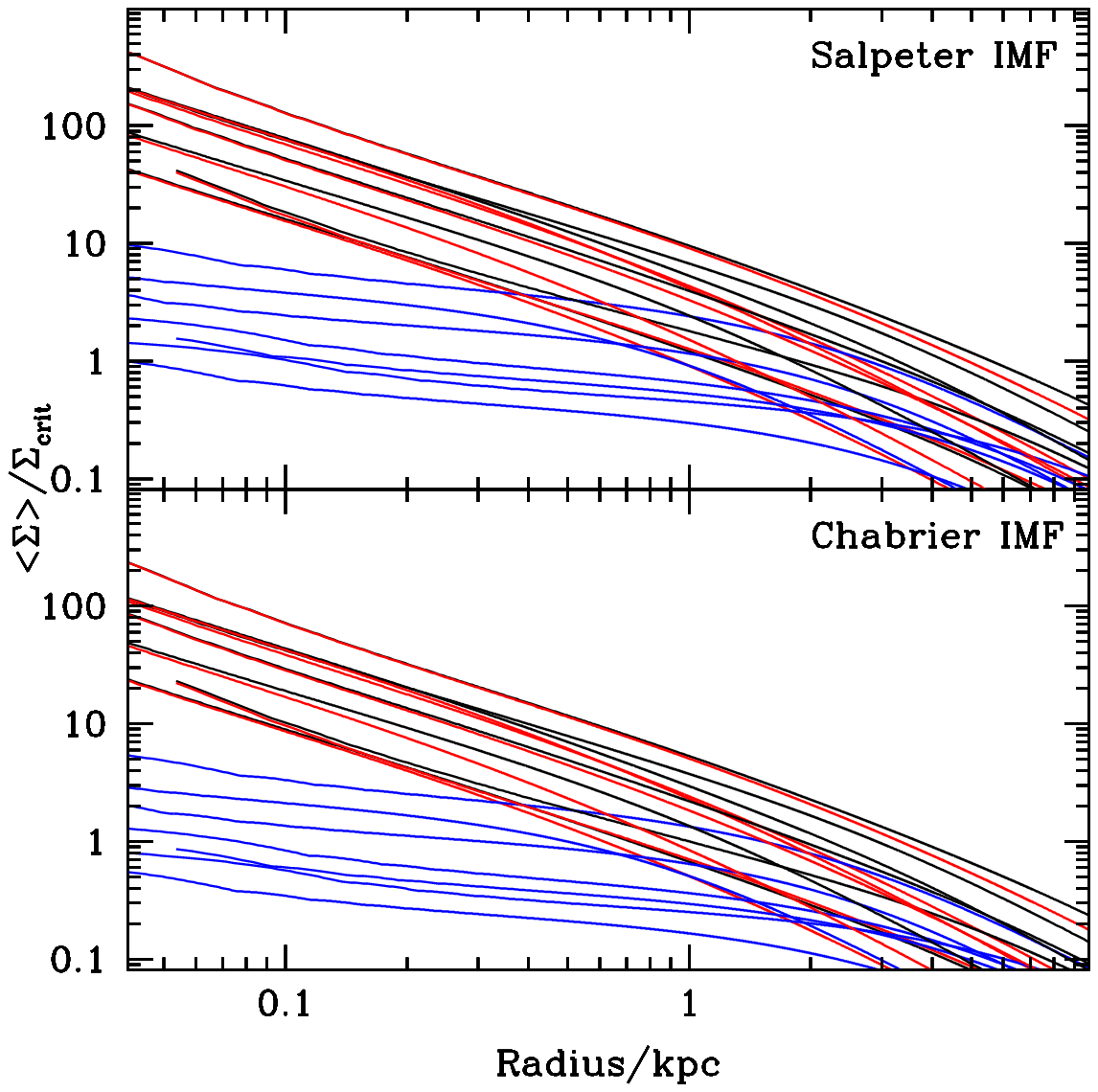}
\centering\includegraphics[width=0.8\linewidth]{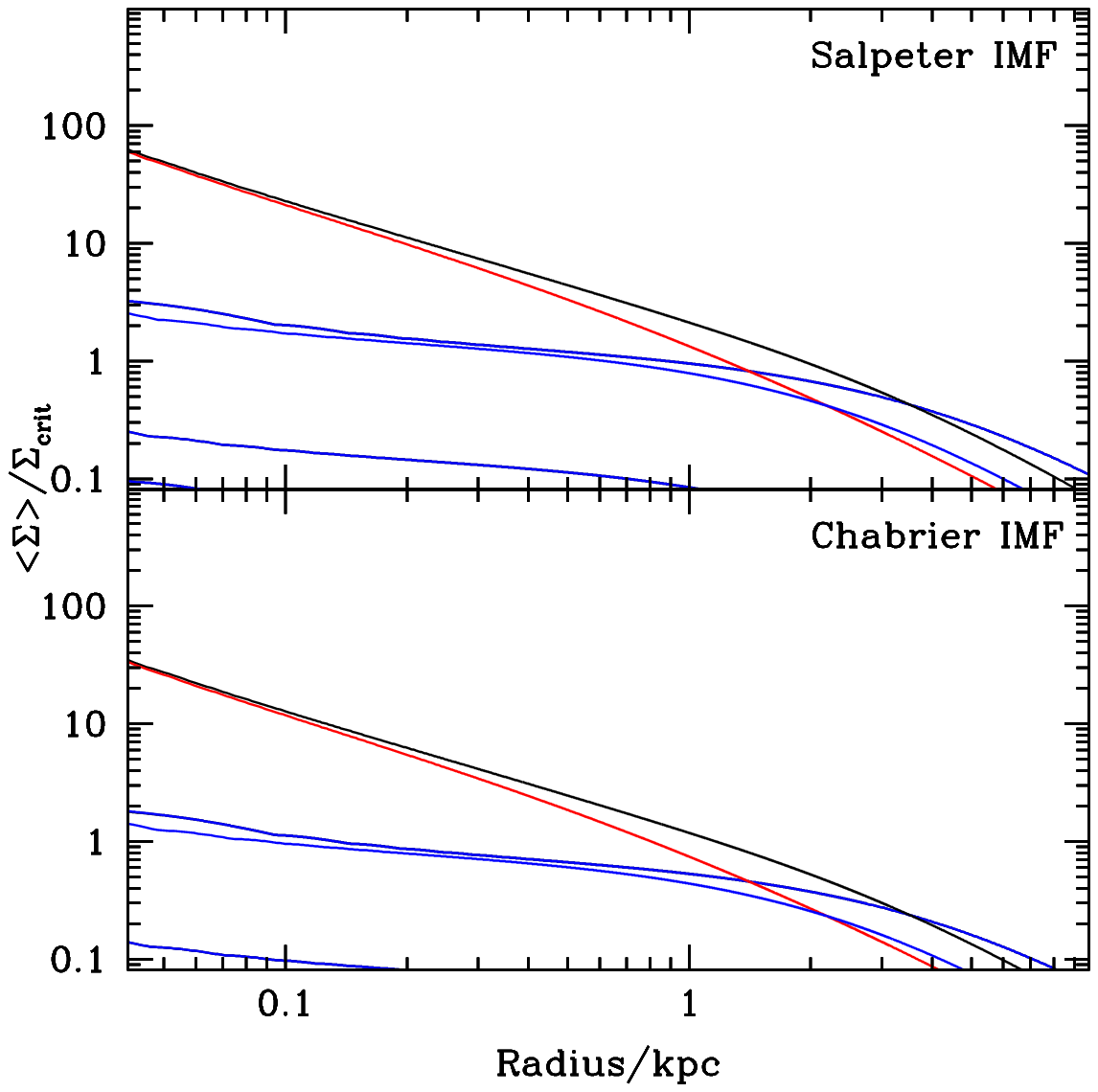}
\caption{Average (enclosed) stellar surface mass density profile of
SWELLS targets, in units of the critical density.  The ``stellar
Einstein radius'' of each lens lies at the point where
$\langle\Sigma\rangle/\Sigma_{\rm crit} \simeq 1$. The black lines
represent the total mass density profile, the red lines represent the
contribution of the bulge, and the blue lines represent that of the
disk. Profiles for both the Salpeter and Chabrier IMF are shown. The
top panel shows the profiles of secure (A) lenses, the middle panel
shows the profile of probable (B) and possible (C) lenses, and the
bottom panel shows the non-lenses (X). Note that the bulge always
dominates the enclosed average mass at small radii, while the disks
start to provide a non-negligible contribution on scales of a kpc or
larger.
\label{fig:kappaprof}}
\end{figure}

\begin{figure}
\centering\includegraphics[width=0.9\linewidth]{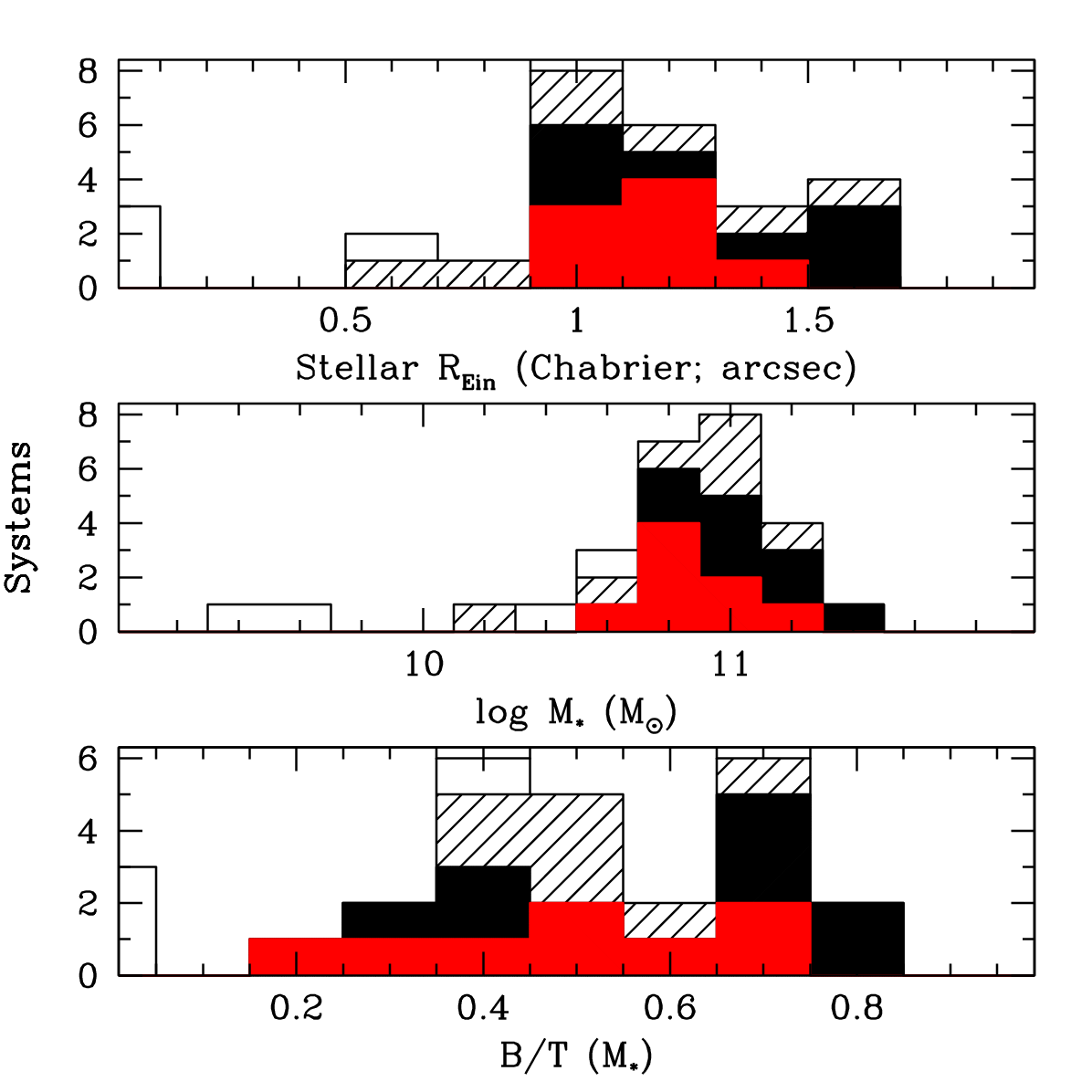}
\caption{Distribution of stellar mass related properties. The top
panel shows the distribution of stellar Einstein radii, i.e. the radii
within which the stellar surface mass density is equal to 1 in
critical units, assuming a Chabrier IMF. This is an approximate lower
limit to the actual Einstein radius, which will include the
contribution of dark matter. The middle panel shows the distribution
of total stellar mass. The bottom panel shows the distribution of
bulge to total stellar mass ratio (the ``bulge fraction'').  As in
Figure~\ref{fig:hist}, the red histogram represents lenses discovered
prior to Cycle~16s; the black solid, hatched and empty histograms
represent lenses targeted in Cycle~16, identified respectively as
secure, probable and possible, and not-lenses. Note how no lens was
confirmed with stellar Einstein radius below $0\farcs94$.
\label{fig:histo2}}
\end{figure}

\subsection{Stellar Mass}
\label{ssec:mass}

Stellar mass estimates for the bulge and disk components of each candidate 
were derived by
comparing multiband photometry with stellar population synthesis models 
\citep{Aug++09}.  The method we employ provides not only stellar
masses but also uncertainties that take into account all the relevant
degeneracies between, e.g., age and metallicity of the stellar
populations.  The dominant systematic uncertainty is the unknown
normalisation of the stellar initial mass function (IMF). Determining
the absolute normalisation of the IMF is one of the goals of the
SWELLS survey (see paper~II). For the moment, we adopt two popular
choices for the IMF, which are believed to bracket the appropriate
range for spiral galaxies \citep{B+d01}. Stellar masses for these
\citet{Cha03} and \citet{Sal55} IMFs are given in
Table~\ref{tab:structure}.

\subsection{Notes on the selection function}
\label{ssec:success}

The overall success rate for confirmation of SWELLS candidates in
Cycle~16s is $42\pm13$\% (8/19) for A-grade and $47\pm14$\% (9/19) for
A+B-grade lenses. As expected, this is somewhat lower than the
confirmation rate for massive early-type galaxies in the SLACS survey,
due to the fact that our galaxies are on average less massive: 
the Einstein radii are smaller and occupy a smaller fraction
of the SDSS fiber. Furthermore, identifying multiple images is
inherently more difficult behind spiral deflectors, due to the
complexity of their surface brightness distribution and the presence
of dust.

We can investigate the selection function more quantitatively by
studying the stellar mass structure of the deflectors and correlations
of its properties with the confirmation rate. Mass and surface mass
density are expected to be the key parameters determining the lensing
strength of a galaxy. We will therefore investigate the dependency of
confirmation rate on several proxies for mass and density.

First, in Figure~\ref{fig:kappaprof} we show the average stellar mass
within circular apertures, expressed in units of the critical density
$\Sigma_{\rm crit}\equiv \frac{c^2 D_{\rm s} }{4\pi G D_{\rm d} D_{\rm
ds} }$, where $D_{\rm s}$ $D_{\rm d}$ and $D_{\rm ds}$ are the angular
diameter distance to the source, to the deflector and between the
deflector and the sources, respectively \citep[see,
e.g.,][]{SEF92,Tre10}.  As shown in Figure~\ref{fig:kappaprof}, the
enclosed average stellar mass density is always dominated by the bulge
in the inner parts, although the disk starts to contribute
significantly beyond 1~kpc. All of the secure systems have deflectors
of uniformly high density, above critical density well beyond one kpc,
even for a Chabrier IMF. The not-lenses span a broader range in
surface mass densities including several systems that are
significantly underdense compared to the grade-A systems. Three out of
four of the X-grade lenses do not have a detectable bulge component
(to our sensitivity limit of M$_*=10^{9}$M$_\odot$).

In a circularly symmetric system the Einstein radius corresponds to
the radius within which the average mass density equals the critical
density \citep{SEF92}. In order to obtain a proxy for the Einstein
radius even when no lensing is detected we define the ``stellar
Einstein radius'' as the radius within which the average stellar mass
density equals the critical density. Note that high resolution
multi-band imaging is necessary to estimate this quantity robustly.
Clearly this is only an approximation for non-circular systems, and
will be a lower limit to the true Einstein radius if dark matter is
present.

We are now in a position to study the distribution of integrated
stellar mass~$M_*$, bulge-to-total stellar mass ratio~$B/T$ and
stellar Einstein radii for the SWELLS targets. As shown in
Figure~\ref{fig:histo2} all three quantities correlate at some level
with confirmation rate. Most of the not-lenses are found at small
$M_*$, small $B/T$, and small stellar Einstein radius. However, the
correlation is not perfect: B and C grade lenses are found even at
M$_*>10^{11}$ and B/T$>0.7$. All three objects with no detected bulge
turned out to be not-lenses. Perhaps the cleanest predictor of
confirmation is stellar Einstein radius: there are no confirmed lenses
below $0\farcs94$ even though there are probable and possible lenses
well above $1''$. For comparison, the stellar Einstein radii for the
confirmed SLACS early-type deflector galaxies are also typically
(90\%) above 0$\farcs$94 although 10\% of the values are in the range
0$\farcs$69-0$\farcs$94. This tail of the distribution extending to
smaller radii is consistent with the idea that multiple images with
smaller separation can be detected in the presence of smooth
early-type deflector galaxies compared to the case of spiral
deflectors. However, the number of SWELLS targets is too small too
reach a statistically significant conclusion.

In conclusion, the selection function of the new SWELLS targets is
complex and comprised of multiple steps. In the first one, we
preselect based on SDSS images edge-on late type galaxies. This is
extremely efficient in producing a sample of lenticular and spiral
galaxies. The second step, confirmation as a lens system, depends on
the properties of the deflector but also has a stochastic
component. Denser systems produce larger Einstein radii, which in turn
make the lensing effect more likely and easier to detect at \hst
resolution. In addition, the relative position on the sky of the
source and deflector, as well as the broad band surface brightness of
the source, ultimately control the presence of detectable multiple
imaging. A third layer of complication is added by the presence of
dust lanes and small scale structure in the surface brightness
distribution of the deflector, which may complicate identification of
multiple images even for a perfectly good strong lensing system. In
the next section we will investigate the effective selection function
of the SWELLS sample by comparing the properties of the SWELLS
deflectors with those of a typical sample of spiral galaxies.


\section{SWELLS galaxies as spiral galaxies}
\label{sec:swellsspirals}

Understanding the selection function of any observational sample is
essential, both in order to generalise results to its parent
population, as well as to compare across samples. In the case of the
SWELLS sample the selection function of the deflectors is sufficiently
complicated that it is extremely hard to compute it accurately from
first principles. We therefore take an alternative approach and
reconstruct the basic features of the selection function {\it a
posteriori} by comparing the properties of the SLACS detectors to
those of a suitable comparison sample of non-lens galaxies. The
comparison sample should be as close as possible in terms of
controllable galaxy parameters, so that significant differences might
be interpreted unambiguously as due to the lensing selection
effect. It is particularly useful to perform this comparison for both
confirmed lenses and not-confirmed lenses as this will help separate
the effects of our SDSS-based pre-selection from those related to the
ability of the deflectors to produce detectable strong lensing events.
In this paper we focus on the properties of the SWELLS deflector
galaxies, and leave for future work the issue of whether they live in
an overdense environment \citep[which turns out not to be the case for
the SLACS lenses, see][]{Aug08,Tre++09}.

Among the quantities that are well-studied and available for large
samples of galaxies, stellar mass and size (and therefore density) are
particularly interesting and likely to influence the lensing selection
function. We therefore choose to focus on the size-stellar mass
correlation as our comparison tool \citep[for a similar analysis of
the SLACS sample, based on the fundamental plane relation
see][]{Tre++06}. 

\begin{figure}
\centering\includegraphics[width=0.99\linewidth]{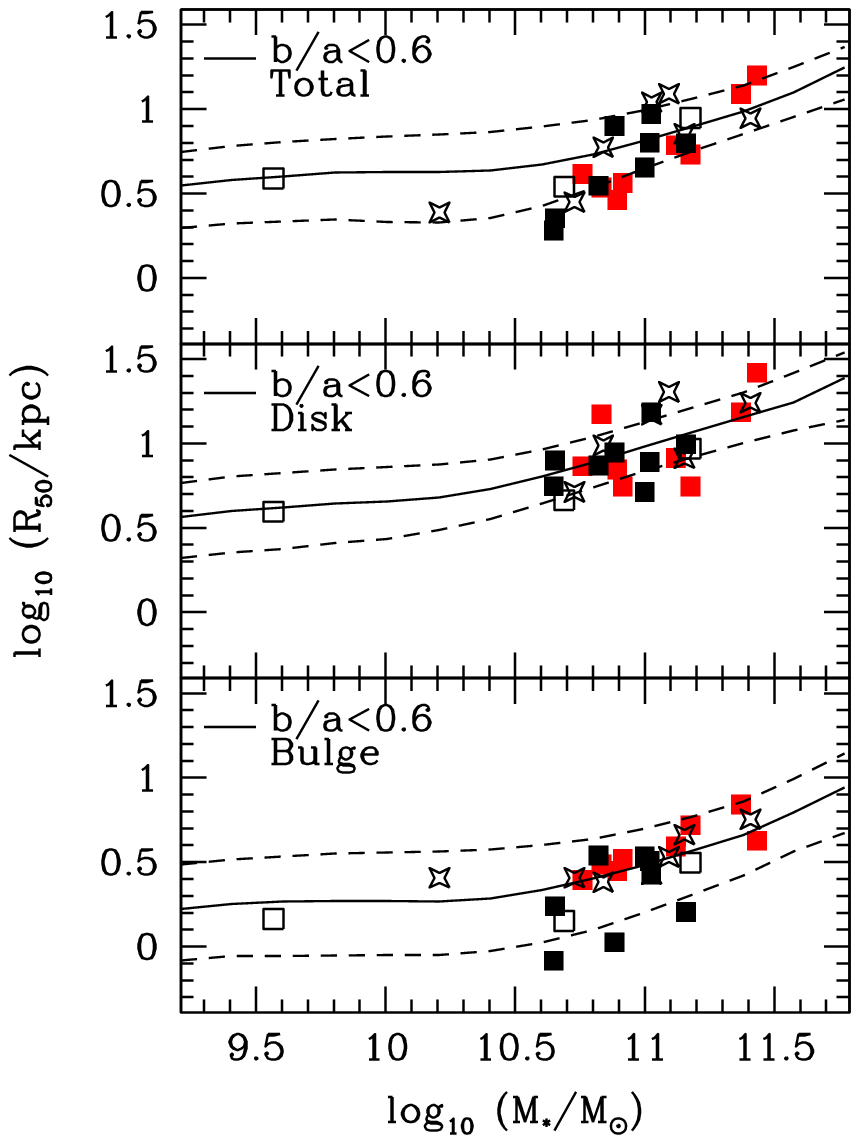}
\caption{The size-mass relation for SWELLS deflector galaxies. 
  The solid and dashed lines represent the
  mean, and 68\% intrinsic scatter, of the size distribution of 
  SDSS galaxies at
  fixed mass with $b/a<0.6$. The points represent the SWELLS galaxies:
  solid squares are secure lenses (A), open stars are probable and
  possible lenses (B and C), while open squares are not lenses
  (X). Red symbols identify pre-Cycle~16s targets, while black symbols
  identify Cycle~16s targets. The top panel shows the total semi-major
  axis effective radius as a function of total stellar mass.  The
  middle panel shows the semi-major axis effective radius of the disk
  component as a function of total stellar mass. The bottom panel
  shows the semi-major axis effective radius of the bulge component as
  a function of total stellar mass. We use the symbol $R_{50}$ to
  identify the semi-major axis effective radius as measured from SDSS
  images, to distinguish it clearly from the circularized half-light
  radius measured from HST images and given in Table~\ref{tab:structure}.
\label{fig:sizemass}}
\end{figure}

A suitable comparison sample for the SWELLS dataset is its parent
sample, the SDSS-DR7 database itself. We adopt the results of the
recent structural analysis performed by L.~Simard and collaborators
(2011, in preparation). These authors fit two-dimensional models to
the SDSS images, using the same parametrisation for bulge and disk as
was adopted in this study. The range in redshift and stellar mass of
their sample is very similar to ours, and the parent sample is the
same. To minimise differences with our sample, we construct the size
mass relation for the SDSS sample limited to $b/a<$0.6. Following
\citet{Dut++10}, we construct the size-mass relationship separately
for the bulge, disk, and total size, adopting the semi-major axis size
for all three components. In contrast to \citet{Dut++10} we do not
apply any selection based on stellar mass or spectral type. To
minimise the impact of measurement biases we perform the comparison by
also using sizes and stellar masses estimated using SDSS images for
our sample of lenses\footnote{The SDSS sizes and masses are in
remarkably good agreement to the \hst determined ones, considering the
difference in resolution. Nevertheless, the small differences are
significant enough that they should be taken into
account. Specifically, there is excellent agreement between the SDSS
and \hst determined stellar mass values, with no significant offset
($0.04\pm0.03$ dex), consistent with the findings of
\citet{Aug++09}. In contrast, the \hst-determined sizes are on average $0.10\pm0.04$ (disk) and $0.17\pm0.06$ dex (bulge) smaller than those determined from SDSS images,
with larger differences for the smaller objects. These differences are
likely due to the higher spatial resolution of \hst and/or a longer
wavelength range used for our \hst fits.}.  The size-mass relation for
SDSS and for our deflector galaxies is shown in
Figure~\ref{fig:sizemass}. The top panel shows the relation between
the total half light radius and total stellar mass. The middle panel
shows the relation between the half light radius of the disk and the
total stellar mass.  The bottom panel shows the relation between the
half light radius of the bulge and the total stellar mass. As a
population, the deflector galaxies are consistent with being drawn
from the SDSS parent population. We quantify this statement by
comparing the vertical offset of each data point with the mean and
standard deviation of the correlation, and computing a $\chi^2$
goodness of fit statistic\footnote{Errors on size can be neglected
since they are much smaller than the intrinsic scatter of the
correlation.}. We find $\chi^2=26,29,9$ respectively for total, disk,
and bulge, with 27 degrees of freedom for total and bulge and 26 for
disk (one galaxy was omitted because of an unphysically small size
returned by the SDSS catalogue, $<1$ kpc). This shows that our
selection of targets {\it at fixed stellar mass} is unbiased with
respect to the parent SDSS population once the $b/a<0.6$ cut is taken
into account. Clearly however, the distribution along the size-mass
relation is very different than that of the parent sample, since our
target list was comprised almost exclusively of systems with mass
above 10$^{10.5}$ M$_\odot$ by design.

It is also interesting to consider whether the distribution of
confirmed lenses is consistent with that of the parent population, or
whether the confirmation rate appears to be a function of, e.g.,
stellar mass density. By comparing the solid (grade A lenses) and open
points (grade B, C, X) in Figure~\ref{fig:sizemass}, it appears that
the solid points tend to lie preferentially somewhat below the open
points. However, the difference is not significant. Even if we
restrict to only the A grade lenses, they still appear to be
consistent with the size mass relation of the SDSS parent population,
at fixed stellar mass ($\chi^2=18,19,7$ for 16 degrees of freedom,
respectively for total, disk, and bulge component).

\section{Finding spiral lens galaxies: lessons learnt and future prospects}
\label{sec:disc}

Finding spiral lens galaxies is more difficult than finding early-type lens
galaxies for two reasons: 1) spiral galaxies tend to be
less massive and less dense than early-type galaxies, and they
therefore produce
smaller image separations; 2) their complex surface brightness
distribution and the presence of dust makes it hard to identify
multiple images if they are present.

For these reasons a crucial element for the success of this search was
the availability of multi-colour high resolution imaging data. 
However, for any given candidate it is
not clear {\it a priori} which band will be most useful.
In general, the SLACS
sources are by selection star-forming and therefore blue \citep[see,
e.g.,][]{New++11}, and stand out the most in the F450W filter. A good
example of this phenomenon is J1117$+$4704, where the quadruply imaged
source near the centre gradually disappears moving to longer
wavelengths. In some cases, however, the presence of dust makes the
infrared data crucial for studying the multiple images. An example of this
phenomenon is J1703$+$2451, where the lensed source is barely visible in
the optical images, while it is obvious in the infrared
one. Interestingly, two of the lenses were confirmed purely on the
basis of $K'$ AO imaging.  It appears that multiband photometry is an
essential ingredient for future studies. At optical wavelengths \hst
is the only option for the time being. With current technology AO is
only feasible for $\sim$1/3 of high latitude targets, and therefore a
combination of AO (when possible) and \hst (when AO is not possible)
observations seems to be the most cost-effective strategy for the infrared.

The other crucial ingredient is angular resolution. This is useful
especially for the detection of compact sources or central
counter-images. Limited resolution is one of the reasons why we do not
confirm any lens with stellar Einstein radius significantly smaller
than one arcsecond.  A good example of this phenomenon is
J0841$+$3824, where the counter-image is clearly visible in $I$-band,
but not at the lower resolution $H$-band, even though this may in part
be due to colour variations. The WFPC2 images obtained in Cycle~16s
significantly under-sample the native \hst PSF. Better-sampled images
with ACS or WFC3-UVIS may result in an improved success rate. Higher
resolution red/infrared images from Next Generation Advanced Optics or
from the James Webb Space Telescope might combine the advantages of
dust removal and angular resolution, even though colour contrast may
still make blue/UV images competitive. Alternatively one may restrict
the search to the systems with larger stellar Einstein radii to
maximize success. However, high-resolution multiband images are needed
to estimate this quantity accurately and therefore it does not appear
a viable strategy for a large sample. Furthermore, one has to be
mindful of the astrophysical implications of such a selection against
objects where dark matter contributes significantly to the central
mass density or where the IMF normalization is higher than normally
assumed. Not to mention that finding lenses with small Einstein Radii
is astrophysically interesting in its own right, in order to probe the
structure of the lower mass galaxies and those with smaller bulges.

In terms of future prospects it seems that another way to improve the
success rate would be to target emission lines in the sources at high
resolution. This would be an efficient way to disentangle light from
the source and deflector, and distinguish without any doubt multiple
images from peculiar features in the surface brightness of the
deflector. The rich filter complement of WFC3 aboard \hst makes it the
most promising instrument for this purpose. Targeting infrared lines
might mitigate the effects of foreground extinction, although with the
present targets only Paschen $\alpha$ is accessible with decent strehl
and the line is typically too faint for present day integral field
spectrographs, even on large 8-10m telescopes. A tunable filter or
custom-made filters will be necessary to pursue this goal.
Alternatively, one could target lines in the radio mapping CO
transitions or HI. ALMA and SKA will have the resolution and
sensitivity to measure emission line maps of the source {\it and}
lens, and will thus also offer the ability to measure kinematics of the
lens and of the source. Molecular gas tends to trace the stars, so any
source that is visibly lensed should also be lensed in CO. Neutral gas
tends to be more extended (typically by a factor of 2 or more) than
stars and molecular gas so this will increase the chances of strong
lensing.


\section{Summary}
\label{sec:summary}

The goal of the SWELLS (Sloan WFC Edge-on Late-type Lens Survey) is to
study the distribution of luminous and dark matter in disk
galaxies. With this goal in mind we have assembled an unprecedented
sample of \NSWELLS\ candidate gravitational lens systems, where the
deflector is a high inclination disk galaxy. The sample is comprised
of \NSESLACS\ systems previously discovered by the SLACS survey for
which we have obtained follow-up multiband high resolution imaging,
and 19 new targets. The new targets were selected from the SDSS
database to have multiple redshifts in the SDSS fiber, as well as
elongated morphology. We present multi-colour imaging of the sample,
as well as structural parameters derived from this data for the whole
sample.  We then study the properties of the deflectors and compare
them to a sample of non-lens galaxies selected from SDSS with the same
criteria as our lens targets. Our main results can be summarised as
follows:

\begin{enumerate}

\item Of the 19 new targets, \NSWELLSA\ are classified as A-grade
(secure) lenses, \NSWELLSB\ is classified as B-grade (probable),
\NSWELLSC\ are classified as C-grade (possible), and \NSWELLSX\ are
classified as X-grade (not lenses). Our global success rate in Cycle~16s 
was $42\pm13\%$ for A-grade lenses. The total number of A-grade
lenses in the SWELLS sample is \NTOTALA.

\item All the deflector galaxies have prominent disks observed at high
inclination, by design. However, the sample spans a broad range of
morphological and spectral types. It includes early-type spirals and
lenticulars as well as later-type spirals, galaxies with strong
H$\alpha$ emission and galaxies with no detectable H$\alpha$
emission. The stellar bulge to disk ratio of confirmed lenses range
between 0.22 and 0.85, with an average of 0.55.

\item All confirmed lenses have stellar masses above 10$^{10.5}$
M$_\odot$ and stellar Einstein radii above $0\farcs94$. None of the
low stellar mass density targets is confirmed to be a strong lens,
consistent with the idea that stellar mass (as opposed to dark matter) 
dominates the mass density profile at the scales of a few kpc.

\item The SWELLS targets and confirmed lenses follow the same
size-stellar mass relation as a comparison sample of high-elongation
($b/a<0.6$) non-lens galaxies selected from the SDSS survey. This
implies that, {\it at given stellar mass}, SWELLS lenses are
consistent with being representative of the overall population of
high-elongation galaxies as identified by SDSS.

\end{enumerate}

We conclude our study by summarising some of the lessons learnt during
this search. Key elements for a successful disk lens search appear to
be high resolution images spanning from the blue to the near infrared,
to exploit colour contrast as well as low dust extinction. The main
remaining limitation is our ability to disentangle light from the
deflector and source. Future studies should increase their success
rate by targeting emission lines in the source at high resolution, via
narrow band imaging or integral field spectroscopy.


\section*{Acknowledgments}
 
We thank Matteo Barnab\`e for useful discussions and suggestions.
TT acknowledges support from the NSF through CAREER award NSF-0642621,
and from the Packard Foundation through a Packard Research Fellowship.
AAD acknowledges financial support from a CITA National Fellowship,
from the National Science Foundation Science and Technology Center
CfAO, managed by UC Santa Cruz under cooperative agreement
No. AST-9876783. AAD and DCK were partially supported by NSF grant AST
08-08133, and by HST grants AR-10664.01-A, HST AR-10965.02-A, and HST
GO-11206.02-A.
PJM was given support by the TABASGO and Kavli Foundations, and the Royal 
Society, in the form of research fellowships.
LVEK acknowledges the support by an NWO-VIDI programme subsidy
(programme number 639.042.505).
This research is supported by NASA through Hubble Space Telescope
programs GO-10587, GO-11202, and GO-11978, and in part by the National
Science Foundation under Grant No. PHY99-07949. and is based on
observations made with the NASA/ESA Hubble Space Telescope and
obtained at the Space Telescope Science Institute, which is operated
by the Association of Universities for Research in Astronomy, Inc.,
under NASA contract NAS 5-26555, and at the W.M. Keck Observatory,
which is operated as a scientific partnership among the California
Institute of Technology, the University of California and the National
Aeronautics and Space Administration. The Observatory was made
possible by the generous financial support of the W.M. Keck
Foundation. The authors wish to recognize and acknowledge the very
significant cultural role and reverence that the summit of Mauna Kea
has always had within the indigenous Hawaiian community.  We are most
fortunate to have the opportunity to conduct observations from this
mountain.
Funding for the SDSS and SDSS-II was provided by the Alfred P. Sloan
Foundation, the Participating Institutions, the National Science
Foundation, the U.S. Department of Energy, the National Aeronautics
and Space Administration, the Japanese Monbukagakusho, the Max Planck
Society, and the Higher Education Funding Council for England. The
SDSS was managed by the Astrophysical Research Consortium for the
Participating Institutions. The SDSS Web Site is http://www.sdss.org/.

\appendix

\section{Spectroscopic pre-selection of SWELLS candidates}
\label{sec:appendix}

The parent sample for the new SWELLS lens search consists of all SDSS
DR7 spectra with either good or marginal plate quality that are
classified by the SDSS spectroscopic reduction pipeline as galaxies.
Noise estimates for all spectra were rescaled on a plate-by-plate
basis to be consistent with the statistics of sky-subtracted sky
spectra on that plate.

The spectroscopic selection algorithm of the original SLACS sample and
our new SWELLS sample both rely upon subtracting a model for the
continuum of the foreground galaxy and scanning the residual spectrum
for higher redshift emission features.  For SLACS, these models were
taken directly from the SDSS redshift pipeline, and consist of the
best-fit combination of 4 PCA-derived eigenspectra.  For the
absorption-dominated systems typical of SLACS, these models provide
essentially noise-limited subtraction of the foreground galaxy
continua.  However, for emission-line foreground spectra, the quality
of continuum subtraction with these models is much worse.  Hence, we
derived new continuum models by (1) generating a new PCA basis with 7
eigenspectra, and (2) masking all common emission features in the
foreground galaxy spectra before fitting each spectrum with this
basis.

The new models were subtracted from the target spectra, and the
residual spectra were scanned automatically for either (1) multiple
emission lines at a single background redshift, or (2) a single
emission line consistent with a marginally resolved [OII]3727 doublet.
Potential [OII] identifications were all checked for the presence of
veto lines that would indicate H$\alpha$, [OIII]5007, or H$\beta$ as a
more likely identification.  The set of spectra identified by the
automated scan was visually inspected by one of us (ASB) and pruned of
obvious false-positive detections related to night-sky subtraction
residuals, broad emission-line wing residuals, and spectra of
generally bad data quality. Neighboring fibers in the multifiber
spectrograph image were also checked to make sure that candidate [OII]
detections were not due to crosstalk from bright emission lines in
adjacent spectra, or to small-scale auroral emission in the night
sky. The final list of potential target spectra was taken as the set
which survived all visual inspections, and for which either (1) each
of three distinct emission lines at the same background redshift was
detected at 4-sigma or greater significance, or (2) a single candidate
background [OII] emission feature was detected at 7.5-sigma or greater
significance.  Candidate [OII] systems were furthermore ranked as "A"
or "B" based upon the appearance of doublet line structure, which can
be marginally resolved at SDSS resolution.

In total, out of almost a million spectra in the SDSS archive, our
search algorithm found more than 200 new high-probability lens
candidates, in addition to those already targeted by previous SLACS
programs (a similar number). The targets for the SWELLS program were
selected from this new pool of candidates as described in the main
text.




\bibliographystyle{apj}

\begin{thebibliography}{51}
\expandafter\ifx\csname natexlab\endcsname\relax\def\natexlab#1{#1}\fi

\bibitem[\protect\citeauthoryear{Auger}{2008}]{Aug08} Auger 
M.~W., 2008, MNRAS, 383, L40 

\bibitem[{{Auger} {et~al.}(2009){Auger}, {Treu}, {Bolton}, {Gavazzi},
  {Koopmans}, {Marshall}, {Bundy}, \& {Moustakas}}]{Aug++09}
{Auger}, M.~W., {Treu}, T., {Bolton}, A.~S., {Gavazzi}, R., {Koopmans},
  L.~V.~E., {Marshall}, P.~J., {Bundy}, K., \& {Moustakas}, L.~A. 2009, \apj,
  705, 1099

\bibitem[{{Auger} {et~al.}(2011){Auger}, {Treu}, {Brewer}, \&
  {Marshall}}]{Aug++11a}
{Auger}, M.~W., {Treu}, T., {Brewer}, B.~J., \& {Marshall}, P.~J. 2011, \mnras,
  411, L6

\bibitem[{{Auger} {et~al.}(2010){Auger}, {Treu}, {Gavazzi}, {Bolton},
  {Koopmans}, \& {Marshall}}]{Aug++10}
{Auger}, M.~W., {Treu}, T., {Gavazzi}, R., {Bolton}, A.~S., {Koopmans},
  L.~V.~E., \& {Marshall}, P.~J. 2010, \apjl, 721, L163

\bibitem[{{Barnabe} {et~al.}(2011){Barnabe}, {Czoske}, {Koopmans}, {Treu}, \&
  {Bolton}}]{Bar++11}
{Barnabe}, M., {Czoske}, O., {Koopmans}, L.~V.~E., {Treu}, T., \& {Bolton},
  A.~S. 2011, MNRAS, in press

\bibitem[{{Bell} \& {de Jong}(2001)}]{B+d01}
{Bell}, E.~F., \& {de Jong}, R.~S. 2001, \apj, 550, 212

\bibitem[{{Bennert} {et~al.}(2011){Bennert}, {Auger}, {Treu}, {Woo}, \&
  {Malkan}}]{Ben++11}
{Bennert}, V.~N., {Auger}, M.~W., {Treu}, T., {Woo}, J., \& {Malkan}, M.~A.
  2011, \apj, 726, 59

\bibitem[\protect\citeauthoryear{Blain, Moller, 
\& Maller}{1999}]{BMM99} Blain A.~W., Moller O., Maller A.~H., 1999, MNRAS, 303, 423 

\bibitem[{{Bolton} {et~al.}(2008){Bolton}, {Burles}, {Koopmans}, {Treu},
  {Gavazzi}, {Moustakas}, {Wayth}, \& {Schlegel}}]{Bol++08}
{Bolton}, A.~S., {Burles}, S., {Koopmans}, L.~V.~E., {Treu}, T., {Gavazzi}, R.,
  {Moustakas}, L.~A., {Wayth}, R., \& {Schlegel}, D.~J. 2008, \apj, 682, 964

\bibitem[{{Bolton} {et~al.}(2006){Bolton}, {Burles}, {Koopmans}, {Treu}, \&
  {Moustakas}}]{Bol++06}
{Bolton}, A.~S., {Burles}, S., {Koopmans}, L.~V.~E., {Treu}, T., \&
  {Moustakas}, L.~A. 2006, \apj, 638, 703

\bibitem[{{Bosma}(1978)}]{Bos78}
{Bosma}, A. 1978, PhD Thesis, Groningen Univ., (1978)

\bibitem[{{Bullock} {et~al.}(2001){Bullock}, {Kolatt}, {Sigad}, {Somerville},
  {Kravtsov}, {Klypin}, {Primack}, \& {Dekel}}]{Bul++01}
{Bullock}, J.~S., {Kolatt}, T.~S., {Sigad}, Y., {Somerville}, R.~S.,
  {Kravtsov}, A.~V., {Klypin}, A.~A., {Primack}, J.~R., \& {Dekel}, A. 2001,
  \mnras, 321, 559

\bibitem[{{Castander} {et~al.}(2006){Castander}, {Treister}, {Maza}, \&
  {Gawiser}}]{Cas++06}
{Castander}, F.~J., {Treister}, E., {Maza}, J., \& {Gawiser}, E. 2006, \apj,
  652, 955

\bibitem[{{Chabrier}(2003)}]{Cha03}
{Chabrier}, G. 2003, \pasp, 115, 763

\bibitem[{{de Blok} {et~al.}(2001){de Blok}, {McGaugh}, \& {Rubin}}]{deBlo++01}
{de Blok}, W.~J.~G., {McGaugh}, S.~S., \& {Rubin}, V.~C. 2001, \aj, 122, 2396

\bibitem[{{de Vaucouleurs}(1948)}]{dev48}
{de Vaucouleurs}, G. 1948, Annales d'Astrophysique, 11, 247

\bibitem[{{Dutton} {et~al.}(2011){Dutton}, {Brewer}, {Marshall}, {Auger},
  {Treu}, {Koo}, {Bolton}, {Holden}, \& {Koopmans}}]{Dut++11b}
{Dutton}, A.~A., {Brewer}, B.~J., {Marshall}, P.~J., {Auger}, M.~W., {Treu},
  T., {Koo}, D.~C., {Bolton}, A.~S., {Holden}, B.~P., \& {Koopmans}, L.~V.~E.
  2011, \mnras, in press, astro-ph/1101.1622

\bibitem[{{Dutton} {et~al.}(2010){Dutton}, {Conroy}, {van den Bosch}, {Simard},
  {Mendel}, {Courteau}, {Dekel}, {More}, \& {Prada}}]{Dut++10}
{Dutton}, A.~A., {Conroy}, C., {van den Bosch}, F.~C., {Simard}, L., {Mendel},
  T., {Courteau}, S., {Dekel}, A., {More}, S., \& {Prada}, F. 2010, \mnras,
  submitted, astro-ph/1012.5859

\bibitem[{{Dutton} {et~al.}(2005){Dutton}, {Courteau}, {de Jong}, \&
  {Carignan}}]{Dut++05}
{Dutton}, A.~A., {Courteau}, S., {de Jong}, R., \& {Carignan}, C. 2005, \apj,
  619, 218

\bibitem[{{Dutton} {et~al.}(2007){Dutton}, {van den Bosch}, {Dekel}, \&
  {Courteau}}]{Dut++07}
{Dutton}, A.~A., {van den Bosch}, F.~C., {Dekel}, A., \& {Courteau}, S. 2007,
  \apj, 654, 27

\bibitem[{{F{\'e}ron} {et~al.}(2009){F{\'e}ron}, {Hjorth}, {McKean}, \&
  {Samsing}}]{Fer++09}
{F{\'e}ron}, C., {Hjorth}, J., {McKean}, J.~P., \& {Samsing}, J. 2009, \apj,
  696, 1319

\bibitem[{{Huchra} {et~al.}(1985){Huchra}, {Gorenstein}, {Kent}, {Shapiro},
  {Smith}, {Horine}, \& {Perley}}]{Huc++85}
{Huchra}, J., {Gorenstein}, M., {Kent}, S., {Shapiro}, I., {Smith}, G.,
  {Horine}, E., \& {Perley}, R. 1985, \aj, 90, 691

\bibitem[{{Jaunsen} \& {Hjorth}(1997)}]{J+H97}
{Jaunsen}, A.~O., \& {Hjorth}, J. 1997, \aap, 317, L39

\bibitem[{{Keeton} \& {Kochanek}(1998)}]{K+K98}
{Keeton}, C.~R., \& {Kochanek}, C.~S. 1998, \apj, 495, 157

\bibitem[{{Klypin} {et~al.}(1999){Klypin}, {Kravtsov}, {Valenzuela}, \&
  {Prada}}]{Kly++99}
{Klypin}, A., {Kravtsov}, A.~V., {Valenzuela}, O., \& {Prada}, F. 1999, \apj,
  522, 82

\bibitem[{{Komatsu} {et~al.}(2009){Komatsu}, {Dunkley}, {Nolta}, {Bennett},
  {Gold}, {Hinshaw}, {Jarosik}, {Larson}, {Limon}, {Page}, {Spergel},
  {Halpern}, {Hill}, {Kogut}, {Meyer}, {Tucker}, {Weiland}, {Wollack}, \&
  {Wright}}]{Kom++09}
{Komatsu}, E., {Dunkley}, J., {Nolta}, M.~R., {Bennett}, C.~L., {Gold}, B.,
  {Hinshaw}, G., {Jarosik}, N., {Larson}, D., {Limon}, M., {Page}, L.,
  {Spergel}, D.~N., {Halpern}, M., {Hill}, R.~S., {Kogut}, A., {Meyer}, S.~S.,
  {Tucker}, G.~S., {Weiland}, J.~L., {Wollack}, E., \& {Wright}, E.~L. 2009,
  \apjs, 180, 330

\bibitem[{{Koopmans} {et~al.}(1998){Koopmans}, {de Bruyn}, \&
  {Jackson}}]{Koo++98}
{Koopmans}, L.~V.~E., {de Bruyn}, A.~G., \& {Jackson}, N. 1998, \mnras, 295,
  534

\bibitem[{{Maller} {et~al.}(1997){Maller}, {Flores}, \& {Primack}}]{MFP97}
{Maller}, A.~H., {Flores}, R.~A., \& {Primack}, J.~R. 1997, \apj, 486, 681

\bibitem[\protect\citeauthoryear{Maller et al.}{2000}]{Mal++00} 
Maller A.~H., Simard L., Guhathakurta P., Hjorth J., Jaunsen A.~O., Flores 
R.~A., Primack J.~R., 2000, ApJ, 533, 194 

\bibitem[{{Marshall} {et~al.}(2009){Marshall}, {Hogg}, {Moustakas},
  {Fassnacht}, {Brada{\v c}}, {Schrabback}, \& {Blandford}}]{Mar++09}
{Marshall}, P.~J., {Hogg}, D.~W., {Moustakas}, L.~A., {Fassnacht}, C.~D.,
  {Brada{\v c}}, M., {Schrabback}, T., \& {Blandford}, R.~D. 2009, \apj, 694,
  924

\bibitem[{{Moore} {et~al.}(1999){Moore}, {Ghigna}, {Governato}, {Lake},
  {Quinn}, {Stadel}, \& {Tozzi}}]{Moo++99}
{Moore}, B., {Ghigna}, S., {Governato}, F., {Lake}, G., {Quinn}, T., {Stadel},
  J., \& {Tozzi}, P. 1999, \apjl, 524, L19

\bibitem[{{More} {et~al.}(2011){More}, {Jahnke}, {More}, {Gallazzi}, {Bell},
  {Barden}, \& {Haeussler}}]{Mor++11}
{More}, A., {Jahnke}, K., {More}, S., {Gallazzi}, A., {Bell}, E.~F., {Barden},
  M., \& {Haeussler}, B. 2011, ArXiv e-prints

\bibitem[{{Navarro} {et~al.}(1997){Navarro}, {Frenk}, \& {White}}]{NFW97}
{Navarro}, J.~F., {Frenk}, C.~S., \& {White}, S.~D.~M. 1997, \apj, 490, 493

\bibitem[{{Newton} {et~al.}(2011){Newton}, {Marshall}, {Treu}, {Auger},
  {Gavazzi}, {Bolton}, {Koopmans}, \& {Moustakas}}]{New++11}
{Newton}, E.~R., {Marshall}, P.~J., {Treu}, T., {Auger}, M.~W., {Gavazzi}, R.,
  {Bolton}, A.~S., {Koopmans}, L.~V.~E., \& {Moustakas}, L.~A. 2011, \apj, in
  press, astro-ph/1104.2608

\bibitem[{{Oke}(1974)}]{Oke74}
{Oke}, J.~B. 1974, \apjs, 27, 21

\bibitem[{{Rubin} {et~al.}(1978){Rubin}, {Thonnard}, \& {Ford}}]{RTF78}
{Rubin}, V.~C., {Thonnard}, N., \& {Ford}, Jr., W.~K. 1978, \apjl, 225, L107

\bibitem[{{Salpeter}(1955)}]{Sal55}
{Salpeter}, E.~E. 1955, \apj, 121, 161

\bibitem[{{Sand} {et~al.}(2008){Sand}, {Treu}, {Ellis}, {Smith}, \&
  {Kneib}}]{San++08}
{Sand}, D.~J., {Treu}, T., {Ellis}, R.~S., {Smith}, G.~P., \& {Kneib}, J.-P.
  2008, \apj, 674, 711

\bibitem[{{Schneider} {et~al.}(1992){Schneider}, {Ehlers}, \& {Falco}}]{SEF92}
{Schneider}, P., {Ehlers}, J., \& {Falco}, E.~E. 1992, Gravitational Lenses,
  XIV, 560 pp.~112 figs..~Springer-Verlag Berlin Heidelberg New York.

\bibitem[{{Spergel} {et~al.}(2007){Spergel}, {Bean}, {Dor{\'e}}, {Nolta},
  {Bennett}, {Dunkley}, {Hinshaw}, {Jarosik}, {Komatsu}, {Page}, {Peiris},
  {Verde}, {Halpern}, {Hill}, {Kogut}, {Limon}, {Meyer}, {Odegard}, {Tucker},
  {Weiland}, {Wollack}, \& {Wright}}]{Spe++07}
{Spergel}, D.~N., {Bean}, R., {Dor{\'e}}, O., {Nolta}, M.~R., {Bennett}, C.~L.,
  {Dunkley}, J., {Hinshaw}, G., {Jarosik}, N., {Komatsu}, E., {Page}, L.,
  {Peiris}, H.~V., {Verde}, L., {Halpern}, M., {Hill}, R.~S., {Kogut}, A.,
  {Limon}, M., {Meyer}, S.~S., {Odegard}, N., {Tucker}, G.~S., {Weiland},
  J.~L., {Wollack}, E., \& {Wright}, E.~L. 2007, \apjs, 170, 377

\bibitem[{{Spiniello} {et~al.}(2011){Spiniello}, {Koopmans}, {Trager},
  {Czoske}, \& {Treu}}]{Spi++11}
{Spiniello}, C., {Koopmans}, L.~V.~E., {Trager}, S.~C., {Czoske}, O., \&
  {Treu}, T. 2011, MNRAS, submitted

\bibitem[{{Swaters} {et~al.}(2003){Swaters}, {Madore}, {van den Bosch}, \&
  {Balcells}}]{Swa++03}
{Swaters}, R.~A., {Madore}, B.~F., {van den Bosch}, F.~C., \& {Balcells}, M.
  2003, \apj, 583, 732

\bibitem[{{Sygnet} {et~al.}(2010){Sygnet}, {Tu}, {Fort}, \&
  {Gavazzi}}]{Syg++10}
{Sygnet}, J.~F., {Tu}, H., {Fort}, B., \& {Gavazzi}, R. 2010, \aap, 517, A25+

\bibitem[{{Treu}(2010)}]{Tre10}
{Treu}, T. 2010, \araa, 48, 87

\bibitem[{{Treu} {et~al.}(2010){Treu}, {Auger}, {Koopmans}, {Gavazzi},
  {Marshall}, \& {Bolton}}]{Tre++10}
{Treu}, T., {Auger}, M.~W., {Koopmans}, L.~V.~E., {Gavazzi}, R., {Marshall},
  P.~J., \& {Bolton}, A.~S. 2010, \apj, 709, 1195

\bibitem[\protect\citeauthoryear{Treu et al.}{2009}]{Tre++09} 
Treu T., Gavazzi R., Gorecki A., Marshall P.~J., Koopmans L.~V.~E., Bolton 
A.~S., Moustakas L.~A., Burles S., 2009, ApJ, 692, 1690 

\bibitem[{{Treu} {et~al.}(2006){Treu}, {Koopmans}, {Bolton}, {Burles}, \&
  {Moustakas}}]{Tre++06}
{Treu}, T., {Koopmans}, L.~V., {Bolton}, A.~S., {Burles}, S., \& {Moustakas},
  L.~A. 2006, \apj, 640, 662

\bibitem[{{Trott} {et~al.}(2010){Trott}, {Treu}, {Koopmans}, \&
  {Webster}}]{Tro++10}
{Trott}, C.~M., {Treu}, T., {Koopmans}, L.~V.~E., \& {Webster}, R.~L. 2010,
  \mnras, 401, 1540

\bibitem[{{van Albada} \& {Sancisi}(1986)}]{A+S86}
{van Albada}, T.~S., \& {Sancisi}, R. 1986, Royal Society of London
  Philosophical Transactions Series A, 320, 447

\bibitem[{{van den Bosch} \& {Swaters}(2001)}]{vdB+S01}
{van den Bosch}, F.~C., \& {Swaters}, R.~A. 2001, \mnras, 325, 1017

\bibitem[{{Winn} {et~al.}(2003){Winn}, {Hall}, \& {Schechter}}]{WHS03}
{Winn}, J.~N., {Hall}, P.~B., \& {Schechter}, P.~L. 2003, \apj, 597, 672

\end{thebibliography}


\label{lastpage}
\bsp

\end{document}